\documentclass[journal,twoside,web]{ieeecolor}
\usepackage{generic}
\usepackage{cite}
\usepackage{amsmath,amssymb,amsfonts}
\usepackage{algorithmic}
\usepackage{graphicx}
\usepackage{textcomp}
\usepackage{comment}
\usepackage{hyperref}

\usepackage{multirow}
\usepackage{tabu}
\usepackage{babel}
\usepackage{booktabs} 
\usepackage[ruled,vlined]{algorithm2e}
\newcommand{\nosemic}{\renewcommand{\@endalgocfline}{\relax}}
\newcommand{\dosemic}{\renewcommand{\@endalgocfline}{\algocf@endline}}

\def\BibTeX{{\rm B\kern-.05em{\sc i\kern-.025em b}\kern-.08em
    T\kern-.1667em\lower.7ex\hbox{E}\kern-.125emX}}
\usepackage{listings}
\usepackage{color}

\definecolor{dkgreen}{rgb}{0,0.6,0}
\definecolor{gray}{rgb}{0.5,0.5,0.5}
\definecolor{mauve}{rgb}{0.58,0,0.82}

\newcommand{\revised}[1]{{\color{black} #1}}
\newcommand{\ie}{\textit{i.e.}}
\newcommand{\eg}{\textit{e.g.}}
\SetKwInput{KwInput}{Input}                
\SetKwInput{KwOutput}{Output}              

\begin{document}
\title{Reinventing 2D Convolutions for 3D Images}
\author{Jiancheng Yang, Xiaoyang Huang, Yi He, Jingwei Xu, Canqian Yang, Guozheng Xu, and Bingbing Ni
	\thanks{This work was supported by National Science Foundation
	of China (U20B2072, 61976137, U1611461). Authors would like to appreciate the Student Innovation Center of SJTU for providing GPUs.}
	\thanks{J. Yang is with Shanghai Jiao Tong University, Shanghai, China, with MoE Key Lab of Articial Intelligence, AI Institute, Shanghai Jiao Tong University, Shanghai, China, and also with Dianei Technology, Shanghai, China (e-mail: jekyll4168@sjtu.edu.cn).}
	\thanks{X. Huang is with Shanghai Jiao Tong University, Shanghai, China, and also with MoE Key Lab of Articial Intelligence, AI Institute, Shanghai Jiao Tong University, Shanghai, China (e-mail: huangxiaoyang@sjtu.edu.cn).}
	\thanks{Yi He is with Dianei Technology, Shanghai, China.}
	\thanks{J. Xu, C. Yang, G. Xu are with Shanghai Jiao Tong University, Shanghai, China.} 
	\thanks{B. Ni is with Shanghai Jiao Tong University, Shanghai, China, with MoE Key Lab of Articial Intelligence, AI Institute, Shanghai Jiao Tong University, Shanghai, China, and also with Huawei Hisilicon, Shanghai, China (e-mail: nibingbing@\{sjtu.edu.cn,hisilicon.com\}). }
	\thanks{J. Yang and X. Huang contributed equally to this article.} 
	\thanks{Corresponding author: B. Ni}}

\maketitle

\begin{abstract}
There have been considerable debates over 2D and 3D representation learning on 3D medical images. 2D approaches could benefit from large-scale 2D pretraining, whereas they are generally weak in capturing large 3D contexts. 3D approaches are natively strong in 3D contexts, however few publicly available 3D medical dataset is large and diverse enough for universal 3D pretraining. Even for hybrid (2D + 3D) approaches, the intrinsic disadvantages within the 2D / 3D parts still exist. In this study, we bridge the gap between 2D and 3D convolutions by reinventing the 2D convolutions. We propose ACS (axial-coronal-sagittal) convolutions to perform natively 3D representation learning, while utilizing the pretrained weights on 2D datasets. In ACS convolutions, 2D convolution kernels are split by channel into three parts, and convoluted separately on the three views (axial, coronal and sagittal) of 3D representations. Theoretically, \textbf{ANY} 2D CNN (ResNet, DenseNet, or DeepLab) is able to be converted into a 3D ACS CNN, with pretrained weight of a same parameter size. Extensive experiments \revised{on several medical benchmarks (including classification, segmentation and detection tasks)} validate the consistent superiority of the pretrained ACS CNNs, over the 2D / 3D CNN counterparts with / without pretraining. Even without pretraining, the ACS convolution can be used as a plug-and-play replacement of standard 3D convolution, with smaller model size and less computation.  
\end{abstract}

\begin{IEEEkeywords}
3D medical images, ACS convolutions, deep learning, 2D-to-3D transfer learning. 
\end{IEEEkeywords}

\section{Introduction}
\begin{figure}[!htb]
	\centering
	\includegraphics[width=\linewidth]{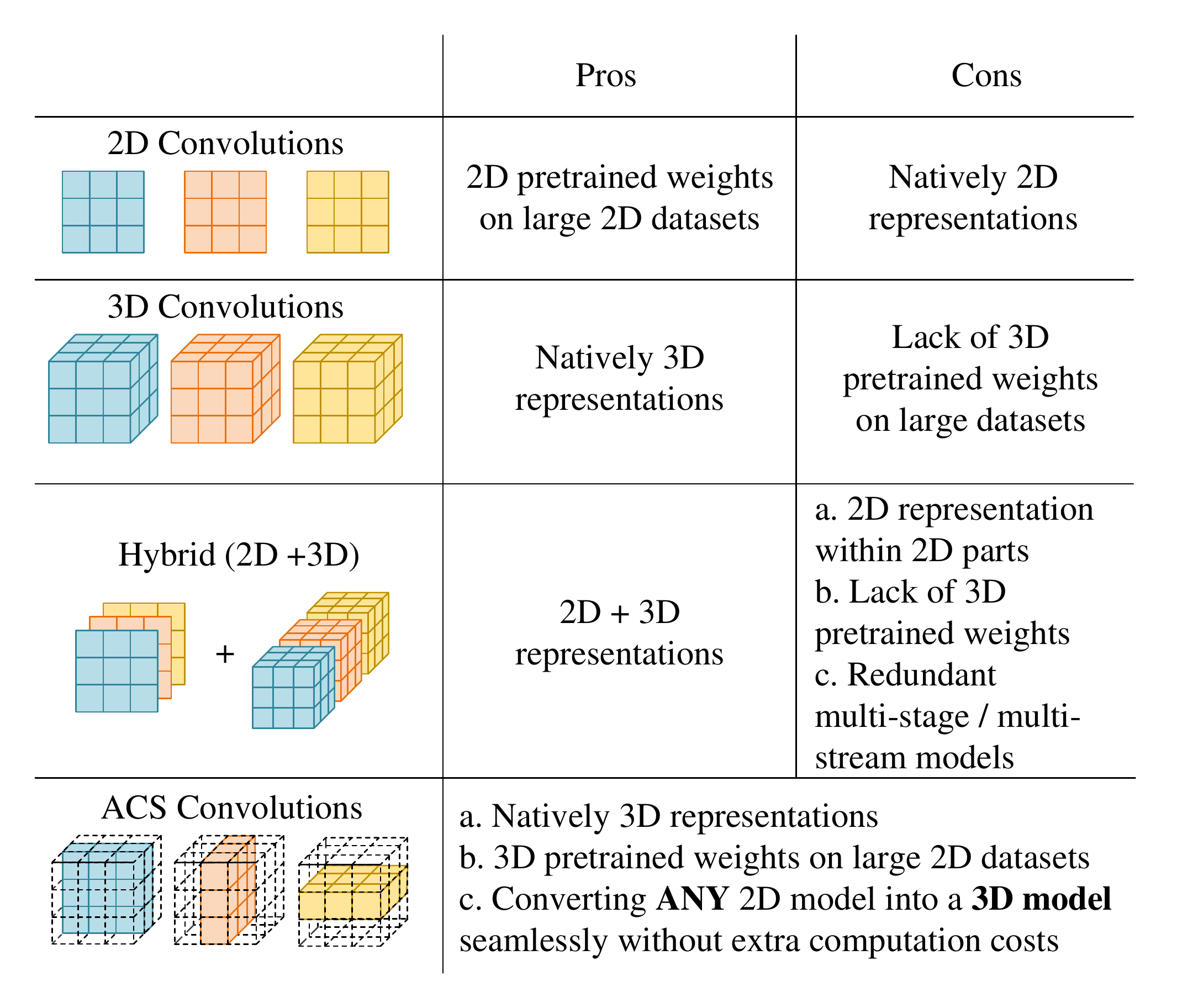} 
	
	\caption{A comparison between the proposed ACS convolutions and prior art on modeling the 3D medical images: pure 2D / 2.5D approaches with 2D convolution kernels, pure 3D approaches with 3D convolution kernels, and hybrid approaches with both 2D and 3D convolution kernels. The ACS convolutions run multiple 2D convolution kernels among the three views (axial, coronal and sagittal). }
	
	\label{fig:acs-illustration}
\end{figure}

\IEEEPARstart{E}{merging} deep learning technology has been dominating the medical image analysis research \cite{litjens2017survey}, in a wide range of data modalities (\eg, ultrasound \cite{droste2019ultrasound}, CT \cite{yan2018deep}, MRI \cite{menze2014multimodal}, X-Ray \cite{wang2017chestx}) and tasks (\eg, classification \cite{gulshan2016development}, segmentation \cite{isensee2018nnu}, detection \cite{yan2019mulan}, registration \cite{balakrishnan2019voxelmorph}). Thanks to contributions from dedicated researchers from academia and industry, there have been much larger medical image datasets than ever before. With large-scale datasets, strong infrastructures and powerful algorithms, numerous challenging problems in medical images seem solvable. However, the data-hungry nature of deep learning limits its applicability in various real-world scenarios with limited annotations. Compared to millions (or even billions) of annotations in natural image datasets, the medical image datasets are not large enough. Especially for 3D medical images, datasets with thousands of supervised training annotations \cite{setio2017validation,simpson2019large} are ``large'' due to imperfect medical annotations \cite{tajbakhsh2020embracing}: hardly-accessible and high dimensional medical data, expensive expert annotators (radiologists / clinicians), and severe class-imbalance issues \cite{yan2019holistic}.

Transfer learning, with pretrained weights from large-scale datasets (\eg, ImageNet \cite{deng2009imagenet}, MS-COCO \cite{lin2014microsoft}), is one of the most important solutions for annotation-efficient deep learning with insufficient data \cite{tajbakhsh2020embracing}. Unfortunately, widely-used pretrained CNNs are developed on 2D datasets, which are non-trivial to transfer to 3D medical images. Prior art on 3D medical images follows either 2D-based approaches or 3D-based approaches (compared in Fig. \ref{fig:acs-illustration}). 2D-based approaches \cite{10.1007/978-3-319-10404-1_65,yu2018recurrent,ni2019elastic} benefit from large-scale pretraining on 2D natural images, while the 2D representation learning are fundamentally weak in large 3D contexts. 3D-based approaches \cite{cciccek20163d,milletari2016v,zhao20183d} learn natively 3D representations. However, few publicly available 3D medical dataset is large and diverse enough for universal 3D pretraining. Therefore, compact network design and sufficient training data are essential for training the 3D networks from scratch. Hybrid (2D + 3D) approaches \cite{li2018h,xia2018bridging,zheng2019new} seem to combine the best of both worlds, nevertheless these ensemble-like approaches do not fundamentally overcome the intrinsic issues of 2D-based and 3D-based approaches. Please refer to Sec. \ref{sec:related-work} for in-depth discussion on these related methods.

There has been considerable debates over 2D and 3D representation learning on 3D medical images: prior studies choose either large-scale 2D pretraining or natively 3D representation learning. This paper presents an alternative to bridge the gap between the 2D and 3D approaches. To solve the intrinsic disadvantages from the 2D convolutions and 3D convolutions in modeling 3D images, we argue that an ideal method should adhere to the following principles:

1) \textbf{\em Natively 3D representation};

2) \textbf{\em 2D weight transferable}: it benefits from 2D pretraining;

3) \textbf{\em ANY model convertible}: it enables any 2D model, including classification, detection and segmentation backbones, to be converted into a 3D one.

These principles cannot be achieved simultaneously with standard 2D convolutions or standard 3D convolutions, which directs us to develop a novel convolution operator. Inspired from the widely-used tri-planar representations of 3D medical images \cite{10.1007/978-3-319-10404-1_65}, we propose \textit{ACS convolutions} satisfying these principles. Instead of explicitly treating the input 3D volumes as three orthogonal 2D planar images \cite{10.1007/978-3-319-10404-1_65} (axial, coronal and sagittal), we operate on the convolution kernels to perform view-based 3D convolutions, via splitting the 2D convolution kernels into three parts by channel. Notably, \textbf{no additional 3D fusion layer} is required to fuse the three-view representations from the 3D convolutions, since they will be seamlessly fused by the subsequent ACS convolution layers (Sec. \ref{sec:method}).

The ACS convolution aims at a generic and plug-and-play replacement of standard 3D convolutions for 3D medical images. Even without pretraining, the ACS convolution is comparable to 3D convolution with a smaller model size and less computation. When pretrained on large 2D datasets, it consistently outperforms 2D / 3D convolution by a large margin. To improve research reproducibility, a PyTorch \cite{paszke2017automatic} implementation of ACS convolution is open-source\footnote{Code is open-source at: \url{https://github.com/M3DV/ACSConv/}.}. Using the provided functions, standard 2D CNNs (\eg, those from PyTorch torchvison package) could be converted into ACS CNNs for 3D images with a single line of code, where 2D pretrained weights could be directly loaded. Compared with 2D models, it introduces no \revised{extra} computation costs, in terms of FLOPs, memory and model size. \revised{The proposed ACS convolutions could be used in various neural networks for diverse tasks; Extensive experiments on several medical benchmarks (including classification, segmentation and detection tasks) validate the consistent effectiveness of the proposed method. }

\section{Related Work} \label{sec:related-work}

\revised{In this section, we first review 2D / 2.5D, 3D and hybrid approaches for 3D medical images, including their advantages and disadvantages. We then discuss the pretraining for 3D medical images by transfer learning and self-supervised learning techniques. Compared with existing 2D / 2.5D / 3D / hybrid approaches, ACS convolution focuses on how to use the existing pretrained 2D networks in a 3D way. Note that the contribution of this study is also orthogonal to pretraining methods. It is possible to pretrain ACS CNNs on 2D images, videos and 3D medical images with supervised or self-supervised learning. This paper uses ACS convolutions with supervised pretraining on 2D natural images.}

\subsection{2D / 2.5D Approaches}

Transfer learning from 2D CNNs, trained on large-scale datasets (\eg, ImageNet \cite{deng2009imagenet}), is a widely-used approach in 3D medical image analysis. To mimic the 3-channel image representation (\ie, RGB), prior studies follow either multi-planar or multi-slice representation of 3D images as 2D inputs. In these studies, pretrained 2D CNNs are usually fine-tuned on the target medical dataset.

Early study \cite{Prasoon2013DeepFL,10.1007/978-3-319-10404-1_65} proposes tri-planar representation of 3D medical images, where three views (axial, coronal and sagittal) from a voxel are regarded as the three channels of 2D input. Although this method is empirically effective, there is a fundamental flaw that the channels are not spatially aligned. More studies follow tri-slice representations \cite{Ding2017AccuratePN,yu2018recurrent,ni2019elastic}, where a center slice together with its two neighbor slices are treated as the three channels. In these representations, the channels are spatially aligned, which conforms to the inductive biases in convolution. There are also studies \cite{yu2018recurrent,perslev2019one} combining both multi-slice and multi-planar approaches, using multi-slice 2D representations in multiple views. The multi-view representations are averaged \cite{yu2018recurrent} or fused by additional networks \cite{perslev2019one}. \revised{Recent work \cite{xia20203d} extracts multi-view information by applying 2D CNNs on rotated and permuted data. Song \textit{et al}.\cite{song2020learning} projects the 3D object boundary surface into a 2D matrix to allow 2D CNN for segmentation.}

Even though these approaches benefit from large-scale 2D pretraining, which is empirically effective in numerous studies \cite{Long2015FullyCN,esteva2017dermatologist,lin2017focal,chen2018encoder}, both multi-slice and multi-planar representation with 2D convolutions are fundamentally weak in capturing large 3D contexts.

\subsection{3D Approaches}

Instead of regarding the 3D spatial information as input channels in 2D approaches, there are numbers of studies using pure 3D convolutions for 3D medical image analysis \cite{cciccek20163d,milletari2016v,kamnitsas2017efficient,dou20173d,zhao20183d}. Compared to limited 3D contexts along certain axis in 2D approaches, the 3D approaches are theoretically capable of capturing arbitrarily large 3D contexts in any axis. Therefore, the 3D approaches are generally better at tasks requiring large 3D contexts, \eg, distinguishing small organs, vessels, and lesions.

However, there are also drawbacks for pure 3D approaches. One of the most important is the lack of large-scale universal 3D pretraining. For this reason, efficient training of 3D networks is a pain point for 3D approaches. Several techniques are introduced to (partially) solve this issue, \eg, deep supervision \cite{dou20173d}, compact network design \cite{zhou2018unet++,zhao20183d}, \revised{\cite{zhang2019light}}. Nevertheless, these techniques are not directly targeting the issue of 3D pretraining. 

A related study of our method is Parallel Separable Convolution (PSC) \cite{Gonda2018ParallelS3} \revised{ and Long-Range Asymmetric Branch (LRAB) \cite{zheng2019hfa}}, which extend pseudo 3D convolution (P3D) \cite{Qiu2017LearningSR} in multiple parallel streams of various directions. Both introduce additional layers apart from 2D convolutions, thereby not all weights could be pretrained. As a comparison, our approach focusing on the use of pretrained 2D weights,  converts whole pretrained networks seamlessly, while keeping same computation as the 2D variants, in terms of FLOPs, memory and parameters (Table \ref{tab:conv-comparison}). \revised{In video analysis, there is work~\cite{li2019collaborative} similar to the proposed Soft-ACS variant.}

\subsection{Hybrid Approaches}

Hybrid approaches are proposed to combine the advantages of both 2D and 3D approaches \cite{li2018h,xia2018bridging,zheng2019new,perslev2019one}. In these studies, 2D pretrained networks with multi-slice inputs, and 3D randomly-initialized networks with volumetric inputs are (jointly or separately) trained for the target tasks.

The hybrid approaches could be mainly categorized into multi-stream and multi-stage approaches. In multi-stream approaches \cite{li2018h,zheng2019new}, 2D networks and 3D networks are designed to perform a same task (\eg, segmentation) in parallel. In multi-stage (\ie, cascade) approaches \cite{xia2018bridging,zheng2019new,perslev2019one}, several 2D networks (and 3D networks) are developed to extract representations from multiple views, and a 3D fusion network is then used to fuse the multi-view representations into 3D representations to peform the target tasks. 

Although empirically effective, the hybrid approaches do not solve the intrinsic disadvantages of 2D and 3D approaches: the 2D parts are still not able to capture large 3D contexts, and the 3D parts still lacks large-scale pretraining. Besides, these ensemble-like methods are generally redundant to deploy.

\subsection{Transfer Learning and Self-Supervised Learning}

Medical annotations require expertise in medicine and radiology, which are thereby expensive to be scalable. For certain rare diseases or novel applications (\eg, predicting response for novel treatment \cite{sun2018radiomics}), the data scale is naturally very small. Transfer learning from large-scale datasets to small-scale datasets is a de-facto paradigm in this case.

Human without any radiological experience could recognize basic anatomy and lesions on 2D and 3D images with limited demonstration. Based on this observation, we believe that transfer learning from universal vision datasets (\eg, ImageNet \cite{deng2009imagenet}, MS-COCO \cite{lin2014microsoft}) should be beneficial for 3D medical image analysis. Although there is literature reporting that universal pretraining is useless for target tasks \cite{he2019rethinking,raghu2019transfusion}, this phenomenon is usually observed when target datasets are large enough. Apart from boosting task performance, the universal pretraining is expected to improve model robustness and uncertainty quantification \cite{pmlr-v97-hendrycks19a}.

Unfortunately, 2D-to-3D transfer learning has not been adequately studied. Research efforts \cite{kamnitsas2017efficient,gibson2018niftynet} have been paid to pretrain natively 3D CNNs on 3D datasets, however few publicly available 3D medical dataset is large and diverse enough for universal pretraining. Prior research explores the transfer learning of 3D CNNs trained on spatio-temporal video datasets \cite{hussein2017risk}. However, there are two kinds of domain shift between video and 3D medical images: 1) natural images vs. medical images, and 2) spatio-temporal data vs. 3D spatial data. The domain shift makes video pretraining \cite{hara2018can} less applicable for 3D medical images. To reduce domain shift, there is research (Med3D \cite{chen2019med3d}) building pretrained 3D models on numbers of 3D medical image datasets. Despite the tremendous effort on collecting data from multiple sources, the data scale of involved 1,000+ training samples is still too much small compared to 1,000,000+ training samples in natural image datasets.

In addition to supervised pretraining, Models Genesis \cite{zhou2019models} explores unsupervised (self-supervised) learning to obtain the pretrained 3D models. Though very impressive, the model performance of up-to-date unsupervised learning is generally not comparable to that of fully supervised learning; even \emph{state-of-the-art} unsupervised / semi-supervised learning techniques \cite{berthelot2019mixmatch,henaff2019data} could not reproduce the model performance using fully supervised training data.

\begin{table}
	
	\caption{A comparison of transfer learning for 3D medical images from various sources.}
	\centering

	\begin{tabular*}{\hsize}{@{}@{\extracolsep{\fill}}lcccc@{}}
		\toprule
		Source & Data Scale & Data Diversity & Supervised & Medical \\
		\midrule
		2D Image  & Very Large & Very Diverse & Y & N \\
		Video \cite{hara2018can}  & Large & Diverse & Y & N \\
		Med3D \cite{chen2019med3d}  & Moderate & Moderate & Y & Y\\
		MG \cite{zhou2019models} & Large & Moderate & N & Y\\
		\bottomrule
	\end{tabular*}

	\label{tab:transfer-comparison}
\end{table}

Table \ref{tab:transfer-comparison} compares the sources of transfer learning for 3D medical images. Compared to transfer learning from video \cite{hara2018can} / Med3D \cite{chen2019med3d} / Models Genesis \cite{zhou2019models}, the key advantage of 2D image pretraining is the \emph{overwhelming} data scale and diversity of datasets. With the ACS convolutions proposed in this study, we are able to develop natively 3D CNNs using 2D pretrained weights. We compare these pretraining approaches in our experiments, and empirically prove the superiority of the proposed ACS convolutions.


\section{ACS Convolutional Neural Networks} \label{sec:method}

\begin{figure*}
	\centering
	\includegraphics[width=0.99\linewidth]{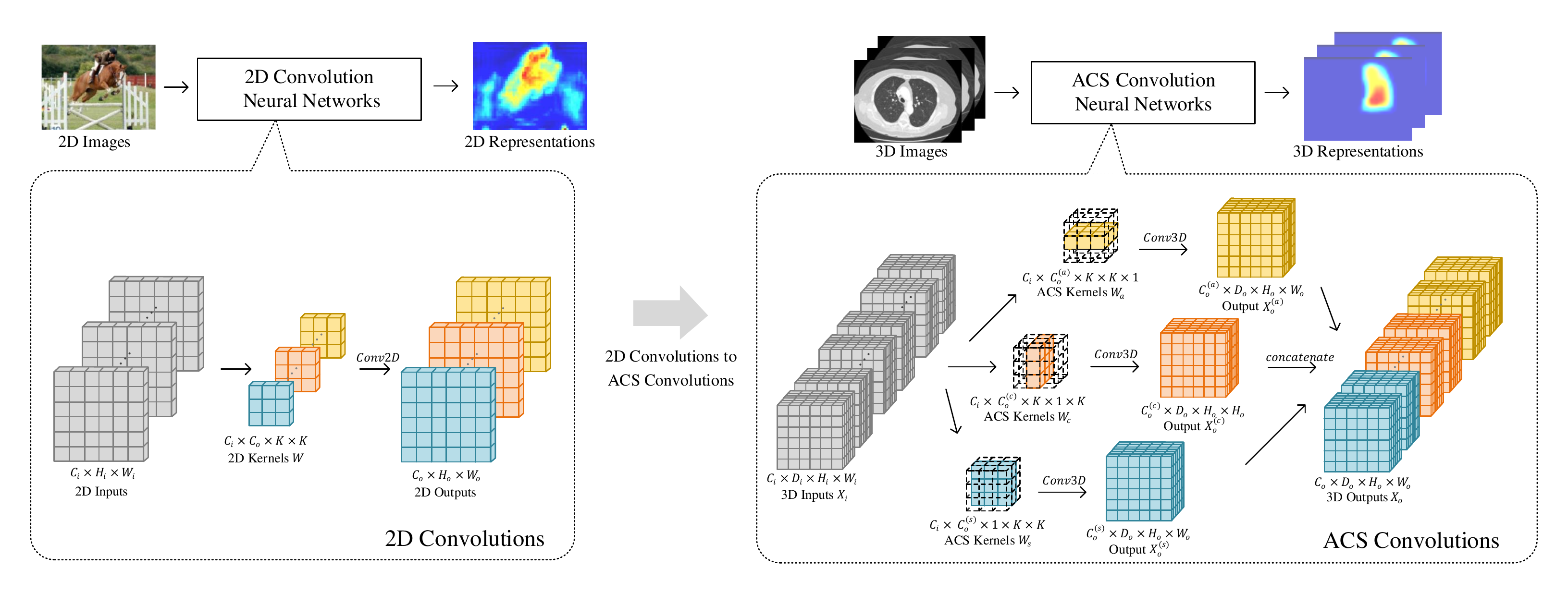} 
	\caption{Illustration of ACS convolutions and 2D-to-ACS model conversion. With a kernel-splitting design, a 2D convolution kernel could be seamlessly transferred into ACS convolution kernels to perform natively 3D representation learning. The ACS convolutions enable \textbf{ANY} 2D model (ResNet \cite{he2016deep}, DenseNet \cite{huang2017densely}, or DeepLab \cite{chen2018encoder}) to be converted into a 3D model.}
	\label{fig:acs-main-figure}
\end{figure*}

\subsection{ACS Convolutions}

Convolution layers capture spatial correlation. Intuitively, the formal difference between 2D and 3D convolutions is the kernel size: the 2D convolutions use 2D kernels (${C_o\times C_i\times K\times K}$) for 2D inputs ($C_i\times H_i\times W_i$), whereas the 3D convolutions use 3D kernels (${C_o\times C_i\times K\times K\times K}$) for 3D inputs ($C_i\times D_i\times H_i\times W_i$), where $C_i$, $C_o$ denote the channels of inputs and outputs, $K$ denotes the kernel size, and $ (D_i\times) H_i\times W_i$ denotes the input size. To transfer the 2D kernels to 3D kernels, there are basically two prior approaches: 1) ``inflate'' the pretrained 2D kernels into 3D kernels size (${K\times K}\rightarrow{K\times K\times K}$), \ie, Inflated 3D (I3D \cite{carreira2017quo}), where the 2D kernels are repeated along an axis and then normalized; 2) unsqueeze the 2D kernels into pseudo 3D kernels on an axis (${K\times K}\rightarrow{1\times K\times K}$), \ie, AH-Net-like \cite{liu20183d}, which could not effectively capture 3D contexts. Note that in both cases, the existing methods assume a specific axis to transfer the 2D kernels. It is meaningful to assign a special axis for spatio-temporal videos, while controversial for 3D medical images. Even for anisotropic medical images, \emph{any view of a 3D image is still a 2D spatial image}.

Based on this observation, we develop ACS (axial-coronal-sagittal) convolutions to learn spatial representations from the axial, coronal and sagittal views. Instead of treating channels of 2D kernels identically \cite{carreira2017quo,liu20183d}, we split the kernels into three parts for extracting 3D spatial information from the axial, coronal and sagittal views. The detailed algorithm of ACS convolutions is shown in the supplementary materials. For simplicity, we introduce ACS convolutions with same padding \revised{as follows}~(Fig. \ref{fig:acs-main-figure}). 

Given a 3D input $\boldsymbol{X_i} \in \mathbb{R}^{C_i\times D_i\times H_i\times W_i}$, we would like to obtain a 3D output $\boldsymbol{X_o} \in \mathbb{R}^{C_o\times D_o\times H_o\times W_o}$, with pretrained / non-pretrained 2D kernels $\boldsymbol{W}\in 
\mathbb{R}^{C_o\times C_i\times K\times K}$. Here, $C_i$ and $C_o$ denote the input and output channels, $D_i\times H_i\times W_i$ and $D_o\times H_o\times W_o$ denote the input and output sizes, $K$ denotes the kernel size. Instead of presenting 3D images into tri-planar 2D images \cite{10.1007/978-3-319-10404-1_65}, we split and reshape the kernels into three parts (named ACS kernels) by the output channel, to obtain the view-based 3D representations for each volume: $\boldsymbol{W_a}\in \mathbb{R}^{C_o^{(a)}\times C_i\times K\times K\times 1}$,  $\boldsymbol{W_c}\in \mathbb{R}^{C_o^{(c)}\times C_i\times  K\times1\times K}$, $\boldsymbol{W_s}\in \mathbb{R}^{C_o^{(s)}\times C_i\times 1\times K\times K}$, where $C_o^{(a)}+C_o^{(c)}+C_o^{(s)}=C_o$. It is theoretically possible to assign an ``optimal axis'' for a 2D kernel; However, considering the feature redundancy in CNNs \cite{han2015deep}, in practice we simply set $C_o^{(a)} \approx C_o^{(c)}  \approx C_o^{(s)} \approx \lfloor C_o/3 \rfloor$. We then compute the axial, coronal and sagittal view-based 3D features via 3D convolutions: 
\begin{equation}
\boldsymbol{X_o^{(a)}} =  \text{Conv3D} (\boldsymbol{X_i},\boldsymbol{W_a})\in \mathbb{R}^{C_o^{(a)}\times D_o\times H_o\times W_o},
\end{equation}
\begin{equation}
\boldsymbol{X_o^{(c)}} =  \text{Conv3D} (\boldsymbol{X_i},\boldsymbol{W_c})\in \mathbb{R}^{C_o^{(c)}\times D_o\times H_o\times W_o},
\end{equation}
\begin{equation}
\boldsymbol{X_o^{(s)}} =  \text{Conv3D} (\boldsymbol{X_i},\boldsymbol{W_s})\in \mathbb{R}^{C_o^{(s)}\times D_o\times H_o\times W_o}.
\end{equation}

The output feature $\boldsymbol{X_o}$ is obtained by concatenating $\boldsymbol{X_o^{(a)}}$, $\boldsymbol{X_o^{(c)}}$ and $\boldsymbol{X_o^{(s)}}$ by the channel axis. It is noteworthy that, \textbf{no 3D fusion layer is required additionally}. The view-based output features will be automatically fused by subsequent convolution layers, without any additional operation, since the convolution kernels are \textbf{not} split by input channel. Thanks to linearity of convolutions, expectation of features from converted ACS convolutions keeps the same as that of 2D ones, thereby no weight rescaling \cite{carreira2017quo} is needed. It is also the prerequisite for the usefulness of 2D pretraining in the converted ACS convolutions. The ACS convolution could be regarded as a special case of 3D convolutions, whose kernels are block sparse.

\begin{table}
	
	\caption{Main operator conversion from 2D CNNs into ACS CNNs. $(\{1,K\})\times K\times K$ and $(1\times)1\times 1$ denote the kernel sizes.}
	
	\centering
	\begin{tabular}{cc}
		\toprule
		2D CNNs & ACS CNNs\\
		\midrule
		Conv2D $K\times K$ & ACSConv $K\times K$\\
		Conv2D $1\times 1$ & Conv3D $1\times 1\times  1$\\
		\{Batch,Group\}Norm2D & \{Batch,Group\}Norm3D \\ 
		\{Max,Avg\}Pool2D $K\times K$ & \{Max,Avg\}Pool3D $\{1,K\}\times K \times K$ \\ 
		\bottomrule
	\end{tabular}
	
	\label{tab:operator}
\end{table}

\revised{The parameter size of ACS convolutions is exactly same as that of 2D convolutions, as the ACS kernels: $\boldsymbol{W_a}$,  $\boldsymbol{W_c}$ and $\boldsymbol{W_s}$ are directly split and reshaped from the 2D kernels $\boldsymbol{W}$, therefore} the proposed method enables ANY 2D model to be converted into a 3D model. Table \ref{tab:operator} lists how operators in 2D CNNs are converted to those in ACS CNNs. Note that the converted models could load the 2D weights directly. 

\subsection{Counterparts and Related Methods} \label{sec-counterpart}

\subsubsection{2D Convolutions}
We include a simple AH-Net-like \cite{liu20183d} 2D counterpart, by replacing all ACS convolutions in ACS CNNs with Conv3D $1\times K\times K$. We name this pseudo 3D counterpart as ``2.5D'' in our experiments, which enables 2D pretrained weight transferring with ease. \revised{The 3D pooling and normalization layers enable 3D context fusion in this case; although insufficient in 3D, we would rather call this variant as ``2.5D''.}

\subsubsection{3D Convolutions} For the 3D counterparts, we replace all convolutions in ACS CNNs with standard 3D convolutions. Various pretraining sources (I3D \cite{carreira2017quo} with 2D images, Med3D \cite{chen2019med3d}, Video \cite{hara2018can}) are included for fair comparison. If there is any difference between the converted 3D models and the pretrained 3D models, we keep the pretrained 3D network architectures to load the pretrained weights. Models Genesis \cite{zhou2019models} uses 3D UNet-based \cite{cciccek20163d,milletari2016v} network architecture. We train the same network from scratch / with its self-supervised pretraining to compare with our models. Moreover, we implement P\textsubscript{3}SC\textsubscript{1} to compare Parallel Separable Convolutions \cite{Gonda2018ParallelS3} with ACS convolutions. The P\textsubscript{3}SC\textsubscript{1} models are trained from scratch due to the lack of pretraining with this method.

\begin{table}[tb]
	\caption{Space and time complexity analysis, for 2D (2.5D), 3D, ACS, Mean-ACS, and Soft-ACS convolutions. \revised{$D \times H \times W$ denotes spatial size of a 3D input, $K$ denotes kernel size (identical kernel for simplicity), and $C_i, C_o$ denote the input and output channels.} Bias terms are not counted in parameters. }
	
	\centering
	\begin{tabular*}{\hsize}{@{}@{\extracolsep{\fill}}cccccc@{}}
		\toprule
		Kernels & FLOPs & Memory & Parameters\\
		\midrule
		Conv2D  & $\mathcal{O}(D  H W  C_o C_i K^2)$ & $DHWC_o$ & $C_o C_i  K^2  $ \\
		Conv3D & $\mathcal{O}(D  H W C_o C_i K^3 )$  &$DHWC_o$ & $C_o C_i  K^3$ \\
		ACSConv & $\mathcal{O}(D  H W C_o C_i K^2)$  & $DHWC_o$&$ C_o C_i  K^2$ \\
		M-ACSConv & $\mathcal{O}(3 DH W C_o C_i K^2)$  &$3DHWC_o$ & $C_o C_i  K^2$ \\
		S-ACSConv &$\mathcal{O}(3 DH W C_o C_i K^2 )$  & $3DHWC_o$& $C_o(C_iK^2+3)$ \\
		\bottomrule
	\end{tabular*}

	\label{tab:conv-comparison}
\end{table}

Table \ref{tab:conv-comparison} compares the time and space complexity of 2D (2.5D), 3D and ACS convolutions. The proposed ACS convolution could be used as a generic and plug-and-play replacement of 3D convolution, with less computation and smaller size. Besides, the ACS convolution enables 2D pretraining. We demonstrate its superiority over the counterparts with extensive experiments (Sec. \ref{sec:result}).

\subsection{ACS Convolution Variants}

Apart from the kernel splitting approach used in the proposed ACS convolutions, there are possible variants to implement the 2D-transferable ACS convolutions.

\subsubsection{Mean-ACS convolutions} Instead of splitting the 2D convolution kernels, we replicate and reshape $\boldsymbol{W}$ into $\boldsymbol{W_a^{'}}\in \mathbb{R}^{C_o\times C_i\times K\times K\times 1}$,  $\boldsymbol{W_c^{'}}\in \mathbb{R}^{C_o\times C_i\times  K\times1\times K}$,
$\boldsymbol{W_s^{'}}\in \mathbb{R}^{C_o\times C_i\times 1\times K\times K}$, and obtain the 3D features $\boldsymbol{X_o^{'(a)}} =  \text{Conv3D} (\boldsymbol{X_i},\boldsymbol{W_a^{'}})$, $\boldsymbol{X_o^{'(c)}} =\text{Conv3D}(\boldsymbol{X_i},\boldsymbol{W_c^{'}})$, $\boldsymbol{X_o^{'(s)}} =  \text{Conv3D} (\boldsymbol{X_i},\boldsymbol{W_s^{'}})$. The output features is
\begin{equation}
\boldsymbol{X_o^{M}} = ( \boldsymbol{X_o^{'(a)}}+\boldsymbol{X_o^{'(c)}}+\boldsymbol{X_o^{'(s)}}) / 3.
\end{equation}
\revised{Note that the replication and reshaping operations for $\boldsymbol{W}$ are conducted on the fly, thereby only one copy of the $\boldsymbol{W}$ are required to be stored for the model.}

\subsubsection{Soft-ACS convolutions} Note that the Mean-ACS convolution uses a symmetric aggregation, thereby it could not distinguish any view-based information. To this regard, we introduce weighted sum of Mean-ACS, \ie, Soft-ACS,
\begin{equation}
\boldsymbol{X_o^{S}} = \alpha^{(a)}\boldsymbol{X_o^{'(a)}}+\alpha^{(c)}\boldsymbol{X_o^{'(c)}}+\alpha^{(s)}\boldsymbol{X_o^{'(s)}},
\end{equation}
where $\alpha^{(a)}, \alpha^{(c)}, \alpha^{(s)}\in\mathbb{R}$ are learnable weights.

In Table \ref{tab:conv-comparison}, we compare the time and space complexity. The two variants are more computationally intensive in terms of FLOPs and memory. Unfortunately, they do not provide significant performance boost empirically. Therefore, we only report the model performance of ACS convolutions in Sec. \ref{sec:result}, and analyze these variants in the supplementary materials.

\section{Experiments} \label{sec:result}

We experiment with the proposed method on a proof-of-concept dataset and medical benchmarks. To fairly compare model performance, we include several counterparts (2.5D/3D/ACS \{Network\} \textbf{r.}/\textbf{p.}) with the same experiment setting, where \textbf{r.} denotes random initialization, and \textbf{p.} denotes pretraining on various sources. We use separate network architectures in different experiments to demonstrate the flexibility and versatility of the proposed method.

\subsection{Proof of Concept: How Does the Pretraining Work?}

\subsubsection{Motivation} We would like to intuitively understand whether the 2D pretraining could be useful for the converted ACS models. For this reason, we design a proof-of-concept experiments with a synthetic dataset, where the ``knowledge'' from the 2D space is guaranteed to be useful in the 3D space. UNet-based models \cite{ronneberger2015u,cciccek20163d} segment the foregrounds, on the dataset consisting of sufficient 2D samples (foreground: circle and square) for pretraining the 2D networks and limited 3D samples (foreground: sphere, cube, cylinder, cone and pyramid) for evaluating the converted ACS networks. Note that the shapes of 2D dataset are exactly the projected single views of 3D volumes (except for triangle), thereby the 2D pretraining is expected to be useful in the 3D segmentation. We quantitatively analyze feature discriminative ability on the 3D dataset \textbf{without} training on it, using a mAUC metric based on Receiver Operating Characteristic (ROC) analysis of final layer features to discriminate the foreground. We then train the UNets and compare the ACS, 2.5D and 3D counterparts. Dice and mIoU averaged on the 5 foreground classes are reported on the 3D dataset. Details of the synthetic dataset, network, training and mAUC metric are provided in the supplementary materials.

\begin{figure}[tb]
	\centering
	\includegraphics[width=\linewidth]{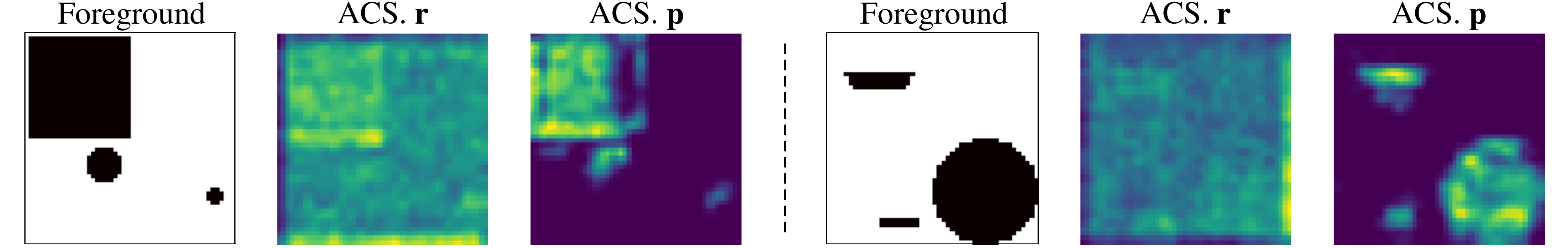} 
	\caption{Examples of features from ACS \textbf{r.} and \textbf{p.} on the 3D dataset \textbf{without} any training on it. ACS \textbf{p.} is pretrained on the 2D dataset. }
	\label{fig:features-without-training}
\end{figure}

\subsubsection{Result Analysis}

We first illustrate two examples of feature maps from ACS \textbf{r.} and ACS \textbf{p.} on the 3D dataset in Fig. \ref{fig:features-without-training}. Even without any training on the target dataset, the features from ACS \textbf{p.} are well aligned with the foreground. As shown in Table \ref{toy-performance}, the mAUC metric of ACS \textbf{p.} is significantly higher than ACS \textbf{r.}, which empirically proves the usefulness of  pretraining for ACS convolutions. Notably, features from ACS \textbf{p.} are even more discriminative than 2.5D \textbf{p.} After training on the 3D dataset, the performance of ACS UNet \textbf{r.} is comparable to 3D UNet \textbf{r.}, and the ACS UNet \textbf{p.} achieves the best performance. The results indicate that ACS convolution is an alternative to 3D convolution with comparable or even better performance, and a smaller model size. 

\begin{table}[tb]
	\caption{Segmentation performance on the proof-of-concept dataset.}
	
	\centering
	\begin{tabular*}{\hsize}{@{}@{\extracolsep{\fill}}lcccc@{}}
		\toprule
		Models & Feature mAUC w/o training & Dice & mIoU & Size  \\
		\midrule
		2.5D UNet \textbf{r.}& 69.0 & 82.2 & 72.5 & 1.6 Mb  \\ 
		2.5D UNet \textbf{p.}& 85.1 & 82.7 & 73.3 & 1.6 Mb  \\
		3D UNet \textbf{r.}& 72.1 & 94.6 & 90.8 & 4.7 Mb  \\ 
		\midrule
		ACS UNet \textbf{r.}& 68.7 & 94.7 & 90.7 & 1.6 Mb  \\ 
		ACS UNet \textbf{p.}& \textbf{88.1}  & \textbf{95.4} & \textbf{92.0} & \textbf{1.6 Mb} \\ 
		\bottomrule
	\end{tabular*}
	
	\label{toy-performance}
\end{table}

\subsection{Lung Nodule Classification and Segmentation}
\label{sec:LIDC}

\begin{table}[tb]
	
	\caption{\revised{LIDC-IDRI~\cite{armato2011lung}} lung nodule segmentation (Dice global) and classification (AUC) performance.}
	
	\centering
	\begin{tabular*}{\hsize}{@{}@{\extracolsep{\fill}}lccccccc@{}}
		\toprule
		Models & Segmentation (Dice) & Classification (AUC) \\
		\midrule
		Models Genesis \cite{zhou2019models} \textbf{r.} & 75.5 & 94.3 \\ 
		Models Genesis \cite{zhou2019models} \textbf{p.} & 75.9 & 94.1 \\ 
		P\textsubscript{3}SC\textsubscript{1} \cite{Gonda2018ParallelS3} \textbf{r.} & 74.3 & 90.9 \\
		\midrule
		2.5D ResNet-18 \textbf{r.} & 68.8 & 89.4 \\ 
		2.5D ResNet-18 \textbf{p.} & 69.8 & 92.0 \\
		3D ResNet-18 \textbf{r.} & 74.7 & 90.3 \\ 
		3D ResNet-18 \textbf{p.} I3D \cite{carreira2017quo} & 75.7 & 91.5 \\ 
		3D ResNet-18 \textbf{p.} Med3D \cite{chen2019med3d}& 74.9 & 90.6 \\ 
		3D ResNet-18 \textbf{p.} Video \cite{hara2018can} & 75.7 & 91.0 \\ 
		\midrule
		ACS ResNet-18 \textbf{r.} & 75.1 & 92.5 \\ 
		ACS ResNet-18 \textbf{p.}  & \textbf{76.5} & \textbf{94.9} \\ 
		\bottomrule
	\end{tabular*}

	\label{tab:lidc-performance}
\end{table}

\begin{figure*}[tb]
	\centering
	\includegraphics[width=0.8\linewidth]{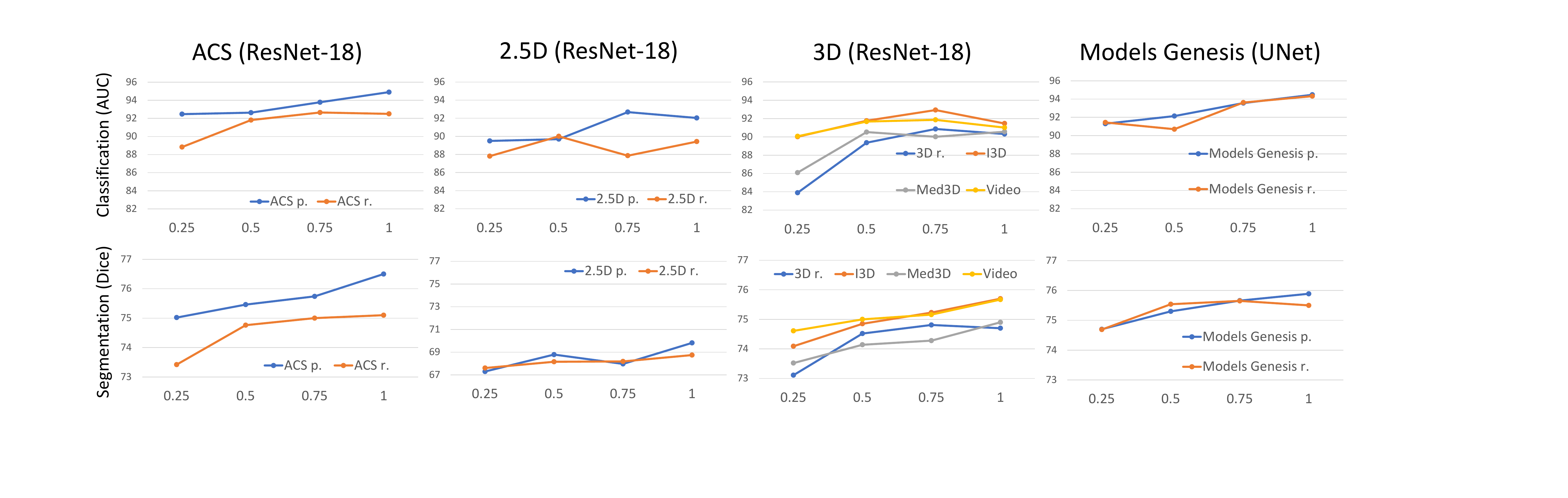} 
	\caption{Performance of ACS, 2.5D, 3D and MG \cite{zhou2019models} on \revised{LIDC-IDRI~\cite{armato2011lung}} lung nodule classification and segmentation vs. the training data scale.}
	\label{fig:lidc_data_scale}
\end{figure*}

\subsubsection{Dataset}
We then validate the effectiveness of the proposed method on a large medical data LIDC-IDRI \cite{armato2011lung}, the largest public lung nodule dataset, for both lung nodule segmentation and malignancy classification task. There are $2,635$ lung nodules annotated by at most $4$ experts, from $1,018$ CT scans. The annotations include pixel-level labelling of the nodules and $5$-level classification of the malignancy, from ``1'' (highly
benign) to ``5'' (highly malignant). For segmentation, we choose one of the up to 4 annotations for all cases. For classification, we take the mode of the annotations as its category. In order to reduce ambiguity, we ignore nodules with level-``3'' (uncertain labelling) and perform binary classification by categorizing the cases with level ``1/2'', ``4/5'' into class 0, 1. It results in a total of 1,633 nodules for classification. We randomly divide the dataset into $4:1$ for training and evaluation, respectively. At training stage we perform data augmentation including random-center cropping, random-axis rotation and flipping.

\subsubsection{Experiment Setting} We compare the ACS models with 2.5D and 3D counterparts with or without pretraining. The pretrained 2.5D / ACS weights are from  PyTorch torchvision package \cite{paszke2017automatic}, trained on ImageNet \cite{deng2009imagenet}. For 3D pretraining, we use the official pretrained models by Med3D \cite{chen2019med3d} and Video\cite{hara2018can}, while I3D \cite{carreira2017quo} weights are transformed from the 2D ImageNet-pretrained weights as well. For PSC \cite{Gonda2018ParallelS3}, we use the configuration of P\textsubscript{3}SC\textsubscript{1}, which resembles our ACS methods. To take advantage of the pretrained weights from Med3D \cite{chen2019med3d} and video \cite{hara2018can} for comparison, all models are adopted a ResNet-18 \cite{he2016deep} architecture, except for Model Genesis \cite{zhou2019models}, since the official pretrained model is based on a 3D UNet \cite{cciccek20163d} architecture. For all model training, we use an Adam optimizer \cite{kingma2014adam} with an initial learning rate of $0.001$ and train the model for $100$ epochs, and delay the learning rate by $0.1$ after $50$ and $75$ epochs. For ResNet-18 backbone, in order to keep higher resolution for output feature maps, we modify the stride of first layer ($7\times7$ stride-2 convolution) into $1$, and remove the first max-pooling. Note that this modification still enables pretraining. A FCN-like \cite{Long2015FullyCN} decoder is applied with progressive upsampling twice. Dice loss with a batch of $8$ is used for segmentation, and binary cross-entropy loss with a batch of $24$ for classification. Dice global and AUC are reported for these two tasks. To demonstrate the flexibility and versatility of ACS convolutions, we also report the results of VGG \cite{Simonyan15} and DenseNet \cite{huang2017densely} in the supplementary materials, which is consistent with the ResNet-18.

To further demonstrate the effectiveness of pretraining, we evaluate models trained with various percentages of training data ($25\%$, $50\%$, $75\%$ and $100\%$) on both segmentation and classification. To maintain the same number of training iterations, the numbers of epoch are increased to $100 / (0.25, 0.5, 0.75, 1)$. The best results among all epochs are reported. We plot the results in Fig. \ref{fig:lidc_data_scale}.

\subsubsection{Result Analysis}

Experiment results are depicted in Table \ref{tab:lidc-performance}. The ACS models consistently outperform all the counterparts by large margins, including 2.5D and 3D models in both random initialization or pretraining settings. P\textsubscript{3}SC\textsubscript{1}\cite{Gonda2018ParallelS3} performance is similar to standard 3D convolution. We observe that the 3D models (ACS, PSC and 3D) generally outperform the 2.5D models, indicating that the usefulness of 3D contexts.  Except for the pretrained 2.5D model on classification task, its superior performance over 3D counterparts may explain the prior art \cite{xie2017transferable,liu2019multi} with 2D networks on this dataset. As for pretraining, the ImageNet \cite{deng2009imagenet} provides significant performance boost (see 2.5D \textbf{p.}, 3D \textbf{p.} I3D \cite{carreira2017quo} and ACS \textbf{p.}), while Med3D \cite{chen2019med3d} brings limited performance boost. We conjecture that it owes to the overwhelming data scale and diversity of 2D image dataset. 

Due to the difference on network architecture (ResNet-based FCN vs. UNet), we experiment with the official code of self-supervised pretrained Models Genesis \cite{zhou2019models} with exactly same setting. Even without pretraining, the segmentation and classification performance of the UNet-based models are strong on this dataset. Despite this, the pretrained ACS model is still better performing. Besides, negative transferring is observed for classification experiments by the Models Genesis \cite{zhou2019models} encoder-only transferring, whereas the ImageNet pretraining consistently improves the performance.
Apart from the superior model performance, the ACS model achieves the best parameter efficiency in our experiments. Take the segmentation task for example, the size of ACS model is 49.8 Mb, compared to 49.8 Mb (2.5D), 142.5 Mb (3D) and 65.4 Mb (MG \cite{zhou2019models}).

As showed in Fig \ref{fig:lidc_data_scale}, models with pretraining consistently outperform those without pretraining under various training data scales. Moreover, when trained with $25\%$ data, the performance gap between \emph{p.} and \emph{n.} reaches the highest, which implies the efficiency of pretraining for limited annotated data. Note that ACS \emph{p.} consistently outperforms all counterparts, no matter how much training data are leveraged.

\subsection{Liver Tumor Segmentation (LiTS) Benchmark}

\subsubsection{Dataset}

We further experiment with our approach on LiTS ~\cite{bilic2019liver}, a challenging 3D medical image segmentation dataest. It consists of $131$ and $70$ enhanced abdominal CT scans for training and testing respectively, to segment the liver and liver tumors. The training annotations are open to public while the test ones are only accessible by online evaluation. The sizes of $x$, $y$ axis are $512$, while the sizes of $z$ axis vary in the range of $[50, 1000]$. We transpose the axes into $z,y,x$ to keep the concept consistent as previously mentioned. For pre-processing, we clip the Hounsfield Unit to $[-200, 250]$ and then normalize to $[0, 1]$. Training data augmentation includes random-center cropping, random-axis flipping and rotation, and random-scale resampling.

\begin{table}[tb]
	\caption{\revised{LiTS~\cite{bilic2019liver}} segmentation performance on lesion and liver. DG: Dice Global. DPC: Dice per Case.}
	
	\centering
	\begin{tabular*}{\hsize}{@{}@{\extracolsep{\fill}}lcccc@{}}
		\toprule
		\multirow{2}{*}{Models}  & \multicolumn{2}{c}{Lesion}&\multicolumn{2}{c}{Liver} \\
		& DG &   DPC  & DG &  DPC \\
		\midrule
		H-DenseUNet \cite{li2018h} & 82.4 & 72.2 & 96.5 & 96.1\\
		Models Genesis \cite{zhou2019models}\footnotemark & - & -&-& 91.13 $\pm$1.51\\
		P\textsubscript{3}SC\textsubscript{1} \cite{Gonda2018ParallelS3} \textbf{r.} & 72.6 & 59.1 & 93.9 & 94.2 \\
		\midrule
		2.5D DeepLab \textbf{r.}  & 72.9 & 56.9 & 93.1 & 92.7 \\ 
		2.5D DeepLab \textbf{p.}  & 73.4 & 60.4 & 92.9 & 92.0 \\ 
		3D DeepLab \textbf{r.}  & 75.5 & 62.6 & 94.8 & 94.8 \\ 
		3D DeepLab \textbf{p.} I3D \cite{carreira2017quo} & 76.5 & 58.2 & 94.1 & 93.4 \\
		3D DeepLab \textbf{p.} Med3D \cite{chen2019med3d} & 67.1 & 53.9 & 92.0 & 93.6 \\ 
		3D DeepLab \textbf{p.} Video \cite{hara2018can} & 66.3 & 56.9 & 92.5 & 93.2 \\ 
		\midrule
		ACS DeepLab \textbf{r.} & 75.3 & 62.4  & 95.0 & 94.9 \\ 
		ACS DeepLab \textbf{p.}   & \textbf{79.1} & \textbf{65.8} & \textbf{96.7} & \textbf{96.2} \\ 
		\bottomrule
	\end{tabular*}

	\label{tab:lits-performance}
\end{table}
\footnotetext{The author only releases the pretrained model on chest CTs, thereby we simply report the evaluation metric provided by the paper. }

\subsubsection{Experiment Setting}

A DeepLabv3+ \cite{chen2018encoder} with a backbone of ResNet-101 \cite{he2016deep} is used in this experiment. The pretrained 2D model is directly obtained from PyTorch torchvision package \cite{paszke2017automatic}. The compared baselines are similar to those in the above LIDC experiment (Sec. \ref{sec:LIDC}). We train all the models for $6000$ epochs. An Adam optimizer \cite{kingma2014adam} is used with an initial learning rate of $0.001$, and we decay the learning rate by $0.1$ after $3000$ and $4500$ epochs. At training stage, we crop the volumes to the size of $64\times224\times 224$. As for testing stage, we crop the volumes to the size of $64\times512\times 512$ and adopt window sliding at a step of $24$ at $z$ axis. Dice global and Dice per case of lesion and liver are reported as standard evaluation on this dataset. 

\subsubsection{Result Analysis}

As shown in Table \ref{tab:lits-performance}, consistent model performance as LIDC experiment (Sec. \ref{sec:LIDC}) can be observed. The pretrained ACS DeepLab achieves better performance than the 2D and 3D counterparts (including self-supervised pretraining \cite{zhou2019models}) by large margins; without pretraining, ACS DeepLab achieves comparable or better performance than 3D DeepLab. According to pretraining results on I3D \cite{carreira2017quo}, Med3D \cite{chen2019med3d} and Video \cite{hara2018can} for 3D DeepLab, negative transferring is observed, probably due to severe domain shift and anisotropy on LiTS dataset. We also report a \emph{state-of-the-art} performance on LiTS dataset using H-DenseUNet\cite{li2018h} as a reference. Note that it adopts a completely different training strategy and network architecture (a cascade of 2D and 3D DenseNet \cite{huang2017densely} models), thereby it is insuitable to compare to our models directly. It is feasible to integrate these orthogonal contributions into ours to improve the model performance.

A key advantage of the proposed ACS convolution is that it enables flexible whole-network conversion together with the pretrained weights, including the task head. In the supplementary materials, we validate the superiority of whole-network weight transferring over encoder-only weight transferring. The previous pretraining methods, \eg, I3D \cite{carreira2017quo}, Med3D \cite{chen2019med3d} and  Video \cite{hara2018can}, hardly take care of the scenarios.

\subsection{Universal Lesion Detection on DeepLesion}

\begin{figure}
	\centering
	\includegraphics[width=\linewidth]{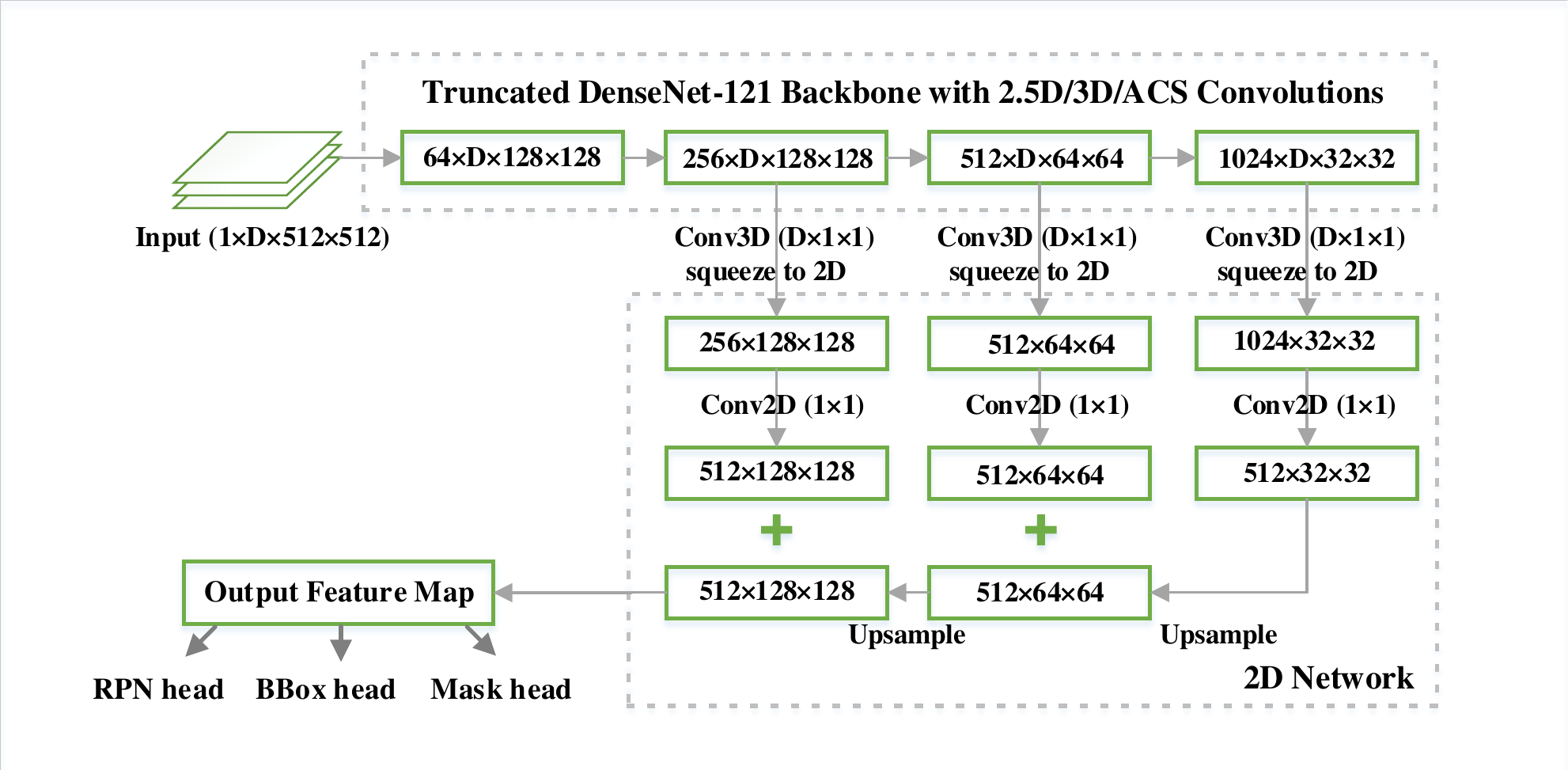} 
	
	\caption{Illustration of the Mask R-CNN \cite{He2017MaskR} for universal lesion detection on DeepLesion \cite{yan2018deep}. The 3D backbone is converted from a truncated DenseNet-121 \cite{huang2017densely,yan2019mulan} with 2.5D/3D/ACS convolutions. }
	
	\label{fig:deeplesion-model}
\end{figure}

\begin{table*}
	
	\caption{Performance on the DeepLesion benchmark \cite{yan2018deep}, in terms of detection sensitivity (Sens, \%) at various false positives (FPs) per image. Note that MULAN \cite{yan2019mulan} uses extra classification tag supervision and an addition Score Refinement Layer (SRL) with tag inputs; We report its performanc under public 171-tag supervision as well as that without SRL. }\label{tab:deeplesion-performance}
	
	\centering
	\begin{tabular*}{\hsize}{@{}@{\extracolsep{\fill}}lccccccccc@{}}
		\toprule
		Methods& Venue &Slices & Sens@0.5 & Sens@1 & Sens@2 & Sens@4 & Sens@8 & Sens@16 & Avg.[0.5,1,2,4]  \\
		\midrule
		3DCE \cite{yan20183d}&MICCAI'18& $\times 27$ & 62.48 & 73.37 &80.70 &85.65 &89.09 &91.06& 75.55 \\
		ULDor \cite{tang2019uldor}&ISBI'19& $\times 1$ & 52.86 & 64.8 & 74.84 & 84.38 & 87.17 & 91.8 & 69.22 \\
		Volumetric Attention \cite{wang2019volumetric}&MICCAI'19& $\times 3$ & 69.10 & 77.90 & 83.80 & - & - & - &- \\
		Improved RetinaNet \cite{zlocha2019improving}&MICCAI'19& $\times 3$ & 72.15 & 80.07 & 86.40 & 90.77 & 94.09 & 96.32 & 82.35\\
		MVP-Net \cite{li2019mvp}&MICCAI'19 & $\times 3$ & 70.01 & 78.77 &84.71 &89.03 &- &- &80.63 \\
		MVP-Net \cite{li2019mvp}&MICCAI'19 & $\times 9$ & 73.83 & 81.82 & 87.60 & 91.30 & - & - & 83.64 \\
		MULAN (Mask R-CNN) \cite{yan2019mulan}&MICCAI'19& $\times 9$  & 76.12 & 83.69 & 88.76 & 92.30 & 94.71 & 95.64 & 85.22  \\
		MULAN (Mask R-CNN) w/o SRL \cite{yan2019mulan}&MICCAI'19& $\times 9$  & -& -&-&-&-&- & 84.22  \\
		\midrule
		
		2.5D Mask R-CNN \textbf{r.} & Ours	&	$\times 3$	&	70.34 	&	79.11 	&	86.16 	&	90.94 	&	94.04 	&	96.01 	&	81.64 		\\
		2.5D Mask R-CNN \textbf{p.}& Ours	&	$\times 3$	&	72.57 	&	79.89 	&	86.80 	&	91.04 	&	94.24 	&	96.32 	&	82.58 		\\
		3D Mask R-CNN \textbf{r.}& Ours	&	$\times 3$	&	63.58 	&	74.63 	&	82.88 	&	88.03 	&	91.39 	&	93.99 	&	77.28 		\\
		3D Mask R-CNN \textbf{p.} I3D \cite{carreira2017quo}& Ours	&	$\times 3$	&	72.01 	&	80.09 	&	86.54 	&	91.29 	&	93.91 	&	95.68 	&	82.48 		\\
		ACS Mask R-CNN \textbf{r.}& Ours	&	$\times 3$	&	72.52 	&	80.85 	&	87.10 	&	91.05 	&	94.39 	&	96.12 	&	82.88 	\\
		ACS Mask R-CNN \textbf{p.}& Ours	&	$\times 3$	&	73.00 	&	81.17 	&	87.05 	&	91.78 	&	94.63 	&	95.48 	&	83.25 		\\
		\midrule
		
		2.5D Mask R-CNN \textbf{r.}& Ours	&	$\times 7$	&	73.37 	&	81.13 	&	86.73 	&	90.96 	&	93.99 	&	95.79 	&	83.05 		\\
		2.5D Mask R-CNN \textbf{p.}& Ours	&	$\times 7$	&	73.66 	&	82.15 	&	87.72 	&	91.38 	&	93.86 	&	95.98 	&	83.73 		\\
		3D Mask R-CNN \textbf{r.}& Ours	&	$\times 7$	&	64.10 	&	75.14 	&	82.13 	&	87.44 	&	91.53 	&	94.22 	&	77.20 	 	\\
		3D Mask R-CNN \textbf{p.} I3D \cite{carreira2017quo}& Ours	&	$\times 7$	&	75.37 	&	83.43 	&	88.68 	&	92.20 	&	94.52 	&	96.07 	&	84.92 	\\
		ACS Mask R-CNN \textbf{r.}& Ours	&	$\times 7$	&	78.01 	&	84.75 	&	88.97 	&	91.76 	&	93.79 	&	95.26 	&	85.87 	\\
		ACS Mask R-CNN \textbf{p.}& Ours	&	$\times 7$	&	\bf78.38 	&	\bf85.39 	&	\bf90.07 	&	\bf93.19 	&	\bf95.18 	&	\bf96.75 	&	\bf86.76 	\\

		\bottomrule
	\end{tabular*}
	
\end{table*}

\begin{figure}
	\centering
	\includegraphics[width=\linewidth]{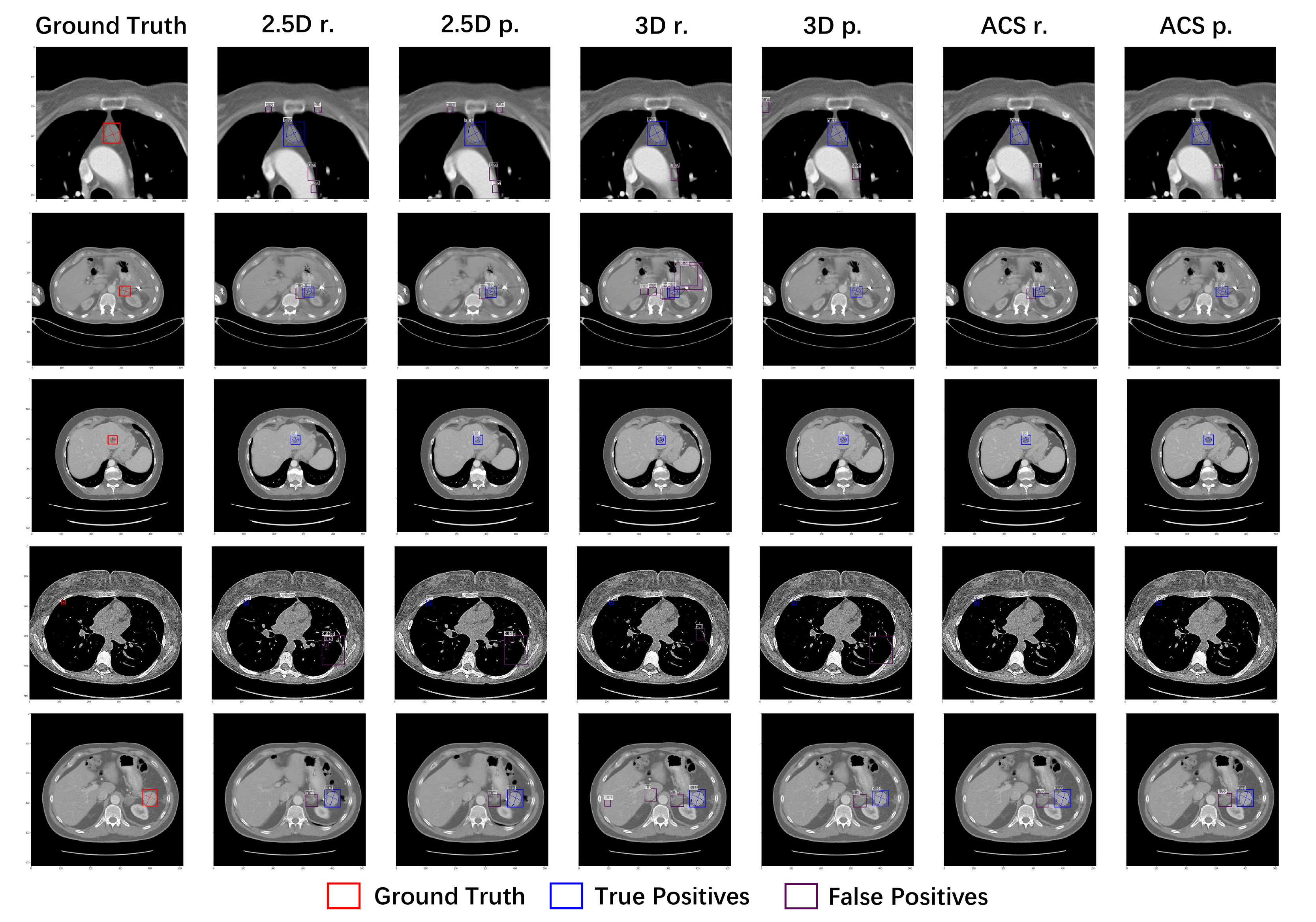} 
	\caption{Illustration of universal lesion detection on the DeepLesion Benchmark. Predicted results with 7-slice inputs are depicted. We also demonstrate the predicted segmentation contours and RECIST diameters in the figure.}
	\label{fig:deeplesion-visualization}
\end{figure}

\subsubsection{Dataset}
DeepLesion dataset \cite{yan2018deep} consists of 32,120 axial CT slices from 10,594 studies of unique patients. There are 1 to 3 lesions in each slice, with totally 32,735 lesions from several organs, whose sizes vary from 0.21 to 342.5mm. RECIST diameter coordinates and bounding boxes were labeled on the key slices, with adjacent slices (above and below 30mm) provided as contextual information. We use GrabCut algorithm to generate weak segmentation ``ground truth'' from weak RECIST labels~\cite{zlocha2019improving}. Hounsfield units of the input are clipped into $[-1024,2050]$ and normalized. The thickness of all slicces is normalized into 2mm.

Data augmentation including horizontal flip, shift, rescaling and rotation is applied during training stage, no test-time augmentation (TTA) is applied. We resize each input slice to $512\times512$ before feeding into the networks. We use official data split (training/validation/test: 70\%/15\%/15\%). To evaluate detection performance, sensitivity at various false positives levels (\ie, FROC analysis) is tuned on the validation set and evaluated on the test set.

\subsubsection{Experiment Setting} As illustrated in Fig. \ref{fig:deeplesion-model}, a detection and instance segmentation network based on Mask R-CNN ~\cite{He2017MaskR} with a same backbone of previous \emph{state-of-the-art} MULAN \cite{yan2019mulan} is developed for universal lesion detection. Since the inputs are 3D slices whereas only 2D key-slice annotations are available on DeepLesion dataset \cite{yan2018deep}, we use a 3D backbone (truncated DenseNet-121 \cite{huang2017densely} converted with
2.5D/3D/ACS convolutions) with a 2D outputs. All 2D $2\times2$ pooling  operators are converted into 3D $1\times2\times2$ pooling. The encoder takes a grey-scale 3D tensor of $1\times D\times 512\times 512$ as input, where $D$ is the length of key slices ($D=3,7$ in this study), and extracts 3D features through three dense blocks. The feature output of each dense blocks is processed by a $D\times 1\times 1$ 3D convolution and then squeezed into a 2D shape. A 2D decoder then combines these features under different resolutions and upsamples the features step by step. The final feature map is fed into RPN head, BBox head and Mask head for detection and instance segmentation supervised by weak RECIST labels. We implement the Mask-RCNN with PyTorch \cite{paszke2017automatic} and MMDetection~\cite{mmdetection}.

\revised{We adopt cross entropy loss for classification and smooth L1 loss for bounding box regression in the RPN head and BBox head, while in the Mask head we adopt the dice loss. These losses weigh equally to form the final loss function. We use an momentum SGD optimizer to train the models for 20 epochs. The learning rate is initialized as 0.02 and multiplied by 0.1 at epoch 10 and 13.}

\subsubsection{Result Analysis} As depicted in Table \ref{tab:deeplesion-performance}, the proposed ACS Mask R-CNN with pretraining significantly outperforms previous \emph{state-of-the-art} 
MULAN \cite{yan2019mulan}. Notably, we use only the detection and RECIST supervision, without additional information beyond the CT images such as tags from medical reports and demographic information. 3D context modeling with 3D and ACS convolutions is proven effective, especially with more input slices. Pretraining consistently improves the model performance for 2.5D, 3D and ACS convolutions. Large performance gap is observed for 3D \textbf{r.} and 3D \revised{\textbf{p.}}, perhaps due to the large model size of 3D convolutions. We also visualize several examples of detection results in Fig. \ref{fig:deeplesion-visualization}. Both 3D context modeling and pretraining reduce the predicted false positives.

\section{Conclusion}
We propose ACS convolution for 3D medical images, as a generic and plug-and-play replacement of standard 3D convolution. It enables pretraining from 2D images, which consistently provides singificant performance boost in our experiments. Even without pretraining, the ACS convolution is comparable or even better than 3D convolution, with smaller model size and less computation. In further study, we will focus on optimal ACS kernel axis assignment and integration with other 2D-to-3D transfer learning operators.


\begin{thebibliography}{10}
\providecommand{\url}[1]{#1}
\csname url@samestyle\endcsname
\providecommand{\newblock}{\relax}
\providecommand{\bibinfo}[2]{#2}
\providecommand{\BIBentrySTDinterwordspacing}{\spaceskip=0pt\relax}
\providecommand{\BIBentryALTinterwordstretchfactor}{4}
\providecommand{\BIBentryALTinterwordspacing}{\spaceskip=\fontdimen2\font plus
\BIBentryALTinterwordstretchfactor\fontdimen3\font minus
  \fontdimen4\font\relax}
\providecommand{\BIBforeignlanguage}[2]{{%
\expandafter\ifx\csname l@#1\endcsname\relax
\typeout{** WARNING: IEEEtran.bst: No hyphenation pattern has been}%
\typeout{** loaded for the language `#1'. Using the pattern for}%
\typeout{** the default language instead.}%
\else
\language=\csname l@#1\endcsname
\fi
#2}}
\providecommand{\BIBdecl}{\relax}
\BIBdecl

\bibitem{litjens2017survey}
G.~Litjens, T.~Kooi, B.~E. Bejnordi, A.~A.~A. Setio, F.~Ciompi, M.~Ghafoorian,
  J.~A. Van Der~Laak, B.~Van~Ginneken, and C.~I. S{\'a}nchez, ``A survey on
  deep learning in medical image analysis,'' \emph{Medical image analysis},
  vol.~42, pp. 60--88, 2017.

\bibitem{droste2019ultrasound}
R.~Droste, Y.~Cai, H.~Sharma, P.~Chatelain, L.~Drukker, A.~T. Papageorghiou,
  and J.~A. Noble, ``Ultrasound image representation learning by modeling
  sonographer visual attention,'' in \emph{International Conference on
  Information Processing in Medical Imaging}.\hskip 1em plus 0.5em minus
  0.4em\relax Springer, 2019, pp. 592--604.

\bibitem{yan2018deep}
K.~Yan, X.~Wang, L.~Lu, L.~Zhang, A.~P. Harrison, M.~Bagheri, and R.~M.
  Summers, ``Deep lesion graphs in the wild: relationship learning and
  organization of significant radiology image findings in a diverse large-scale
  lesion database,'' in \emph{CVPR}, 2018, pp. 9261--9270.

\bibitem{menze2014multimodal}
B.~H. Menze, A.~Jakab, S.~Bauer, J.~Kalpathy-Cramer, K.~Farahani, J.~Kirby,
  Y.~Burren, N.~Porz, J.~Slotboom, R.~Wiest \emph{et~al.}, ``The multimodal
  brain tumor image segmentation benchmark (brats),'' \emph{IEEE transactions
  on medical imaging}, vol.~34, no.~10, pp. 1993--2024, 2014.

\bibitem{wang2017chestx}
X.~Wang, Y.~Peng, L.~Lu, Z.~Lu, M.~Bagheri, and R.~M. Summers, ``Chestx-ray8:
  Hospital-scale chest x-ray database and benchmarks on weakly-supervised
  classification and localization of common thorax diseases,'' in \emph{CVPR},
  2017, pp. 2097--2106.

\bibitem{gulshan2016development}
V.~Gulshan, L.~Peng, M.~Coram, M.~C. Stumpe, D.~Wu, A.~Narayanaswamy,
  S.~Venugopalan, K.~Widner, T.~Madams, J.~Cuadros \emph{et~al.}, ``Development
  and validation of a deep learning algorithm for detection of diabetic
  retinopathy in retinal fundus photographs,'' \emph{Jama}, vol. 316, no.~22,
  pp. 2402--2410, 2016.

\bibitem{isensee2018nnu}
F.~Isensee, J.~Petersen, A.~Klein, D.~Zimmerer, P.~F. Jaeger, S.~Kohl,
  J.~Wasserthal, G.~Koehler, T.~Norajitra, S.~Wirkert \emph{et~al.}, ``nnu-net:
  Self-adapting framework for u-net-based medical image segmentation,''
  \emph{arXiv preprint arXiv:1809.10486}, 2018.

\bibitem{yan2019mulan}
K.~Yan, Y.~Tang, Y.~Peng, V.~Sandfort, M.~Bagheri, Z.~Lu, and R.~M. Summers,
  ``Mulan: Multitask universal lesion analysis network for joint lesion
  detection, tagging, and segmentation,'' in \emph{MICCAI}.\hskip 1em plus
  0.5em minus 0.4em\relax Springer, 2019, pp. 194--202.

\bibitem{balakrishnan2019voxelmorph}
G.~Balakrishnan, A.~Zhao, M.~R. Sabuncu, J.~Guttag, and A.~V. Dalca,
  ``Voxelmorph: a learning framework for deformable medical image
  registration,'' \emph{IEEE transactions on medical imaging}, 2019.

\bibitem{setio2017validation}
A.~A.~A. Setio, A.~Traverso, T.~De~Bel, M.~S. Berens, C.~van~den Bogaard,
  P.~Cerello, H.~Chen, Q.~Dou, M.~E. Fantacci, B.~Geurts \emph{et~al.},
  ``Validation, comparison, and combination of algorithms for automatic
  detection of pulmonary nodules in computed tomography images: the luna16
  challenge,'' \emph{Medical image analysis}, vol.~42, pp. 1--13, 2017.

\bibitem{simpson2019large}
A.~L. Simpson, M.~Antonelli, S.~Bakas, M.~Bilello, K.~Farahani, B.~van
  Ginneken, A.~Kopp-Schneider, B.~A. Landman, G.~Litjens, B.~Menze
  \emph{et~al.}, ``A large annotated medical image dataset for the development
  and evaluation of segmentation algorithms,'' \emph{arXiv preprint
  arXiv:1902.09063}, 2019.

\bibitem{tajbakhsh2020embracing}
N.~Tajbakhsh, L.~Jeyaseelan, Q.~Li, J.~Chiang, Z.~Wu, and X.~Ding, ``Embracing
  imperfect datasets: A review of deep learning solutions for medical image
  segmentation.'' \emph{Medical image analysis}, vol.~63, p. 101693, 2020.

\bibitem{yan2019holistic}
K.~Yan, Y.~Peng, V.~Sandfort, M.~Bagheri, Z.~Lu, and R.~M. Summers, ``Holistic
  and comprehensive annotation of clinically significant findings on diverse ct
  images: Learning from radiology reports and label ontology,'' in \emph{CVPR},
  2019, pp. 8523--8532.

\bibitem{deng2009imagenet}
J.~Deng, W.~Dong, R.~Socher, L.-J. Li, K.~Li, and L.~Fei-Fei, ``Imagenet: A
  large-scale hierarchical image database,'' in \emph{CVPR}.\hskip 1em plus
  0.5em minus 0.4em\relax Ieee, 2009, pp. 248--255.

\bibitem{lin2014microsoft}
T.-Y. Lin, M.~Maire, S.~Belongie, J.~Hays, P.~Perona, D.~Ramanan,
  P.~Doll{\'a}r, and C.~L. Zitnick, ``Microsoft coco: Common objects in
  context,'' in \emph{ECCV}.\hskip 1em plus 0.5em minus 0.4em\relax Springer,
  2014, pp. 740--755.

\bibitem{10.1007/978-3-319-10404-1_65}
H.~R. Roth, L.~Lu, A.~Seff, K.~M. Cherry, J.~Hoffman, S.~Wang, J.~Liu,
  E.~Turkbey, and R.~M. Summers, ``A new 2.5 d representation for lymph node
  detection using random sets of deep convolutional neural network
  observations,'' in \emph{MICCAI}.\hskip 1em plus 0.5em minus 0.4em\relax
  Springer, 2014, pp. 520--527.

\bibitem{yu2018recurrent}
Q.~Yu, L.~Xie, Y.~Wang, Y.~Zhou, E.~K. Fishman, and A.~L. Yuille, ``Recurrent
  saliency transformation network: Incorporating multi-stage visual cues for
  small organ segmentation,'' in \emph{CVPR}, 2018, pp. 8280--8289.

\bibitem{ni2019elastic}
T.~Ni, L.~Xie, H.~Zheng, E.~K. Fishman, and A.~L. Yuille, ``Elastic boundary
  projection for 3d medical image segmentation,'' in \emph{CVPR}, 2019, pp.
  2109--2118.

\bibitem{cciccek20163d}
{\"O}.~{\c{C}}i{\c{c}}ek, A.~Abdulkadir, S.~S. Lienkamp, T.~Brox, and
  O.~Ronneberger, ``3d u-net: learning dense volumetric segmentation from
  sparse annotation,'' in \emph{MICCAI}.\hskip 1em plus 0.5em minus 0.4em\relax
  Springer, 2016, pp. 424--432.

\bibitem{milletari2016v}
F.~Milletari, N.~Navab, and S.-A. Ahmadi, ``V-net: Fully convolutional neural
  networks for volumetric medical image segmentation,'' in \emph{3DV}.\hskip
  1em plus 0.5em minus 0.4em\relax IEEE, 2016, pp. 565--571.

\bibitem{zhao20183d}
W.~Zhao, J.~Yang, Y.~Sun, C.~Li, W.~Wu, L.~Jin, Z.~Yang, B.~Ni, P.~Gao, P.~Wang
  \emph{et~al.}, ``3d deep learning from ct scans predicts tumor invasiveness
  of subcentimeter pulmonary adenocarcinomas,'' \emph{Cancer research},
  vol.~78, no.~24, pp. 6881--6889, 2018.

\bibitem{li2018h}
X.~Li, H.~Chen, X.~Qi, Q.~Dou, C.-W. Fu, and P.-A. Heng, ``H-denseunet: hybrid
  densely connected unet for liver and tumor segmentation from ct volumes,''
  \emph{IEEE transactions on medical imaging}, vol.~37, no.~12, pp. 2663--2674,
  2018.

\bibitem{xia2018bridging}
Y.~Xia, L.~Xie, F.~Liu, Z.~Zhu, E.~K. Fishman, and A.~L. Yuille, ``Bridging the
  gap between 2d and 3d organ segmentation with volumetric fusion net,'' in
  \emph{MICCAI}.\hskip 1em plus 0.5em minus 0.4em\relax Springer, 2018, pp.
  445--453.

\bibitem{zheng2019new}
H.~Zheng, Y.~Zhang, L.~Yang, P.~Liang, Z.~Zhao, C.~Wang, and D.~Z. Chen, ``A
  new ensemble learning framework for 3d biomedical image segmentation,'' in
  \emph{AAAI}, vol.~33, 2019, pp. 5909--5916.

\bibitem{paszke2017automatic}
A.~Paszke, S.~Gross, S.~Chintala, G.~Chanan, E.~Yang, Z.~DeVito, Z.~Lin,
  A.~Desmaison, L.~Antiga, and A.~Lerer, ``Automatic differentiation in
  {PyTorch},'' in \emph{NIPS Autodiff Workshop}, 2017.

\bibitem{Prasoon2013DeepFL}
A.~Prasoon, K.~Petersen, C.~Igel, F.~Lauze, E.~Dam, and M.~Nielsen, ``Deep
  feature learning for knee cartilage segmentation using a triplanar
  convolutional neural network,'' \emph{MICCAI}, vol. 16 Pt 2, pp. 246--53,
  2013.

\bibitem{Ding2017AccuratePN}
J.~Ding, A.~Li, Z.~Hu, and L.~Wang, ``Accurate pulmonary nodule detection in
  computed tomography images using deep convolutional neural networks,'' in
  \emph{MICCAI}, 2017.

\bibitem{perslev2019one}
M.~Perslev, E.~B. Dam, A.~Pai, and C.~Igel, ``One network to segment them all:
  A general, lightweight system for accurate 3d medical image segmentation,''
  in \emph{MICCAI}.\hskip 1em plus 0.5em minus 0.4em\relax Springer, 2019, pp.
  30--38.

\bibitem{xia20203d}
Y.~Xia, F.~Liu, D.~Yang, J.~Cai, L.~Yu, Z.~Zhu, D.~Xu, A.~Yuille, and H.~Roth,
  ``3d semi-supervised learning with uncertainty-aware multi-view
  co-training,'' in \emph{The IEEE Winter Conference on Applications of
  Computer Vision}, 2020, pp. 3646--3655.

\bibitem{song2020learning}
Y.~Song, Z.~Yu, T.~Zhou, J.~Y.-C. Teoh, B.~Lei, K.-S. Choi, and J.~Qin,
  ``Learning 3d features with 2d cnns via surface projection for ct volume
  segmentation,'' in \emph{International Conference on Medical Image Computing
  and Computer-Assisted Intervention}.\hskip 1em plus 0.5em minus 0.4em\relax
  Springer, 2020, pp. 176--186.

\bibitem{Long2015FullyCN}
J.~Long, E.~Shelhamer, and T.~Darrell, ``Fully convolutional networks for
  semantic segmentation,'' \emph{CVPR}, pp. 3431--3440, 2015.

\bibitem{esteva2017dermatologist}
A.~Esteva, B.~Kuprel, R.~A. Novoa, J.~Ko, S.~M. Swetter, H.~M. Blau, and
  S.~Thrun, ``Dermatologist-level classification of skin cancer with deep
  neural networks,'' \emph{Nature}, vol. 542, no. 7639, p. 115, 2017.

\bibitem{lin2017focal}
T.-Y. Lin, P.~Goyal, R.~Girshick, K.~He, and P.~Doll{\'a}r, ``Focal loss for
  dense object detection,'' in \emph{ICCV}, 2017, pp. 2980--2988.

\bibitem{chen2018encoder}
L.-C. Chen, Y.~Zhu, G.~Papandreou, F.~Schroff, and H.~Adam, ``Encoder-decoder
  with atrous separable convolution for semantic image segmentation,'' in
  \emph{ECCV}, 2018, pp. 801--818.

\bibitem{kamnitsas2017efficient}
K.~Kamnitsas, C.~Ledig, V.~F. Newcombe, J.~P. Simpson, A.~D. Kane, D.~K. Menon,
  D.~Rueckert, and B.~Glocker, ``Efficient multi-scale 3d cnn with fully
  connected crf for accurate brain lesion segmentation,'' \emph{Medical image
  analysis}, vol.~36, pp. 61--78, 2017.

\bibitem{dou20173d}
Q.~Dou, L.~Yu, H.~Chen, Y.~Jin, X.~Yang, J.~Qin, and P.-A. Heng, ``3d deeply
  supervised network for automated segmentation of volumetric medical images,''
  \emph{Medical image analysis}, vol.~41, pp. 40--54, 2017.

\bibitem{zhou2018unet++}
Z.~Zhou, M.~M.~R. Siddiquee, N.~Tajbakhsh, and J.~Liang, ``Unet++: A nested
  u-net architecture for medical image segmentation,'' in \emph{Deep Learning
  in Medical Image Analysis and Multimodal Learning for Clinical Decision
  Support}.\hskip 1em plus 0.5em minus 0.4em\relax Springer, 2018, pp. 3--11.

\bibitem{zhang2019light}
J.~Zhang, Y.~Xie, P.~Zhang, H.~Chen, Y.~Xia, and C.~Shen, ``Light-weight hybrid
  convolutional network for liver tumor segmentation.'' in \emph{IJCAI}, 2019,
  pp. 4271--4277.

\bibitem{Gonda2018ParallelS3}
F.~Gonda, D.~Wei, T.~Parag, and H.~Pfister, ``Parallel separable 3d convolution
  for video and volumetric data understanding,'' in \emph{BMVC}, 2018.

\bibitem{zheng2019hfa}
H.~Zheng, L.~Yang, J.~Han, Y.~Zhang, P.~Liang, Z.~Zhao, C.~Wang, and D.~Z.
  Chen, ``Hfa-net: 3d cardiovascular image segmentation with asymmetrical
  pooling and content-aware fusion,'' in \emph{International Conference on
  Medical Image Computing and Computer-Assisted Intervention}.\hskip 1em plus
  0.5em minus 0.4em\relax Springer, 2019, pp. 759--767.

\bibitem{Qiu2017LearningSR}
Z.~Qiu, T.~Yao, and T.~Mei, ``Learning spatio-temporal representation with
  pseudo-3d residual networks,'' \emph{ICCV}, pp. 5534--5542, 2017.

\bibitem{li2019collaborative}
C.~Li, Q.~Zhong, D.~Xie, and S.~Pu, ``Collaborative spatiotemporal feature
  learning for video action recognition,'' in \emph{Proceedings of the IEEE
  Conference on Computer Vision and Pattern Recognition}, 2019, pp. 7872--7881.

\bibitem{sun2018radiomics}
R.~Sun, E.~J. Limkin, M.~Vakalopoulou, L.~Dercle, S.~Champiat, S.~R. Han,
  L.~Verlingue, D.~Brandao, A.~Lancia, S.~Ammari \emph{et~al.}, ``A radiomics
  approach to assess tumour-infiltrating cd8 cells and response to anti-pd-1 or
  anti-pd-l1 immunotherapy: an imaging biomarker, retrospective multicohort
  study,'' \emph{The Lancet Oncology}, vol.~19, no.~9, pp. 1180--1191, 2018.

\bibitem{he2019rethinking}
K.~He, R.~Girshick, and P.~Doll{\'a}r, ``Rethinking imagenet pre-training,'' in
  \emph{CVPR}, 2019, pp. 4918--4927.

\bibitem{raghu2019transfusion}
M.~Raghu, C.~Zhang, J.~Kleinberg, and S.~Bengio, ``Transfusion: Understanding
  transfer learning with applications to medical imaging,'' in \emph{NeurIPS},
  2019.

\bibitem{pmlr-v97-hendrycks19a}
D.~Hendrycks, K.~Lee, and M.~Mazeika, ``Using pre-training can improve model
  robustness and uncertainty,'' in \emph{ICML}, ser. Proceedings of Machine
  Learning Research, K.~Chaudhuri and R.~Salakhutdinov, Eds., vol.~97.\hskip
  1em plus 0.5em minus 0.4em\relax Long Beach, California, USA: PMLR, 09--15
  Jun 2019, pp. 2712--2721.

\bibitem{gibson2018niftynet}
E.~Gibson, W.~Li, C.~Sudre, L.~Fidon, D.~I. Shakir, G.~Wang, Z.~Eaton-Rosen,
  R.~Gray, T.~Doel, Y.~Hu \emph{et~al.}, ``Niftynet: a deep-learning platform
  for medical imaging,'' \emph{Computer methods and programs in biomedicine},
  vol. 158, pp. 113--122, 2018.

\bibitem{hussein2017risk}
S.~Hussein, K.~Cao, Q.~Song, and U.~Bagci, ``Risk stratification of lung
  nodules using 3d cnn-based multi-task learning,'' in \emph{International
  conference on information processing in medical imaging}.\hskip 1em plus
  0.5em minus 0.4em\relax Springer, 2017, pp. 249--260.

\bibitem{hara2018can}
K.~Hara, H.~Kataoka, and Y.~Satoh, ``Can spatiotemporal 3d cnns retrace the
  history of 2d cnns and imagenet?'' in \emph{Proceedings of the IEEE
  conference on Computer Vision and Pattern Recognition}, 2018, pp. 6546--6555.

\bibitem{chen2019med3d}
S.~Chen, K.~Ma, and Y.~Zheng, ``Med3d: Transfer learning for 3d medical image
  analysis,'' \emph{arXiv preprint arXiv:1904.00625}, 2019.

\bibitem{zhou2019models}
Z.~Zhou, V.~Sodha, M.~M.~R. Siddiquee, R.~Feng, N.~Tajbakhsh, M.~B. Gotway, and
  J.~Liang, ``Models genesis: Generic autodidactic models for 3d medical image
  analysis,'' in \emph{MICCAI}.\hskip 1em plus 0.5em minus 0.4em\relax
  Springer, 2019, pp. 384--393.

\bibitem{berthelot2019mixmatch}
D.~Berthelot, N.~Carlini, I.~Goodfellow, N.~Papernot, A.~Oliver, and C.~Raffel,
  ``Mixmatch: A holistic approach to semi-supervised learning,'' \emph{arXiv
  preprint arXiv:1905.02249}, 2019.

\bibitem{henaff2019data}
O.~J. H{\'e}naff, A.~Razavi, C.~Doersch, S.~Eslami, and A.~v.~d. Oord,
  ``Data-efficient image recognition with contrastive predictive coding,''
  \emph{arXiv preprint arXiv:1905.09272}, 2019.

\bibitem{he2016deep}
K.~He, X.~Zhang, S.~Ren, and J.~Sun, ``Deep residual learning for image
  recognition,'' in \emph{CVPR}, 2016, pp. 770--778.

\bibitem{huang2017densely}
G.~Huang, Z.~Liu, L.~Van Der~Maaten, and K.~Q. Weinberger, ``Densely connected
  convolutional networks,'' in \emph{CVPR}, 2017, pp. 4700--4708.

\bibitem{carreira2017quo}
J.~Carreira and A.~Zisserman, ``Quo vadis, action recognition? a new model and
  the kinetics dataset,'' in \emph{CVPR}, 2017, pp. 6299--6308.

\bibitem{liu20183d}
S.~Liu, D.~Xu, S.~K. Zhou, O.~Pauly, S.~Grbic, T.~Mertelmeier, J.~Wicklein,
  A.~Jerebko, W.~Cai, and D.~Comaniciu, ``3d anisotropic hybrid network:
  Transferring convolutional features from 2d images to 3d anisotropic
  volumes,'' in \emph{MICCAI}.\hskip 1em plus 0.5em minus 0.4em\relax Springer,
  2018, pp. 851--858.

\bibitem{han2015deep}
S.~Han, H.~Mao, and W.~J. Dally, ``Deep compression: Compressing deep neural
  networks with pruning, trained quantization and huffman coding,'' \emph{arXiv
  preprint arXiv:1510.00149}, 2015.

\bibitem{ronneberger2015u}
O.~Ronneberger, P.~Fischer, and T.~Brox, ``U-net: Convolutional networks for
  biomedical image segmentation,'' in \emph{International Conference on Medical
  image computing and computer-assisted intervention}.\hskip 1em plus 0.5em
  minus 0.4em\relax Springer, 2015, pp. 234--241.

\bibitem{armato2011lung}
S.~G. Armato~III, G.~McLennan, L.~Bidaut, M.~F. McNitt-Gray, C.~R. Meyer, A.~P.
  Reeves, B.~Zhao, D.~R. Aberle, C.~I. Henschke, E.~A. Hoffman \emph{et~al.},
  ``The lung image database consortium (lidc) and image database resource
  initiative (idri): a completed reference database of lung nodules on ct
  scans,'' \emph{Medical physics}, vol.~38, no.~2, pp. 915--931, 2011.

\bibitem{kingma2014adam}
D.~P. Kingma and J.~Ba, ``Adam: A method for stochastic optimization,'' in
  \emph{ICLR}, 2014.

\bibitem{Simonyan15}
K.~Simonyan and A.~Zisserman, ``Very deep convolutional networks for
  large-scale image recognition,'' in \emph{ICLR}, 2015.

\bibitem{xie2017transferable}
Y.~Xie, Y.~Xia, J.~Zhang, D.~D. Feng, M.~Fulham, and W.~Cai, ``Transferable
  multi-model ensemble for benign-malignant lung nodule classification on chest
  ct,'' in \emph{MICCAI}.\hskip 1em plus 0.5em minus 0.4em\relax Springer,
  2017, pp. 656--664.

\bibitem{liu2019multi}
L.~Liu, Q.~Dou, H.~Chen, J.~Qin, and P.-A. Heng, ``Multi-task deep model with
  margin ranking loss for lung nodule analysis,'' \emph{IEEE transactions on
  medical imaging}, 2019.

\bibitem{bilic2019liver}
P.~Bilic, P.~F. Christ, E.~Vorontsov, G.~Chlebus, H.~Chen, Q.~Dou, C.-W. Fu,
  X.~Han, P.-A. Heng, J.~Hesser \emph{et~al.}, ``The liver tumor segmentation
  benchmark (lits),'' \emph{arXiv preprint arXiv:1901.04056}, 2019.

\bibitem{He2017MaskR}
K.~He, G.~Gkioxari, P.~Doll{\'a}r, and R.~B. Girshick, ``Mask r-cnn,''
  \emph{ICCV}, pp. 2980--2988, 2017.

\bibitem{yan20183d}
K.~Yan, M.~Bagheri, and R.~M. Summers, ``3d context enhanced region-based
  convolutional neural network for end-to-end lesion detection,'' in
  \emph{MICCAI}.\hskip 1em plus 0.5em minus 0.4em\relax Springer, 2018, pp.
  511--519.

\bibitem{tang2019uldor}
Y.-B. Tang, K.~Yan, Y.-X. Tang, J.~Liu, J.~Xiao, and R.~M. Summers, ``Uldor: a
  universal lesion detector for ct scans with pseudo masks and hard negative
  example mining,'' in \emph{ISBI}.\hskip 1em plus 0.5em minus 0.4em\relax
  IEEE, 2019, pp. 833--836.

\bibitem{wang2019volumetric}
X.~Wang, S.~Han, Y.~Chen, D.~Gao, and N.~Vasconcelos, ``Volumetric attention
  for 3d medical image segmentation and detection,'' in \emph{MICCAI}.\hskip
  1em plus 0.5em minus 0.4em\relax Springer, 2019, pp. 175--184.

\bibitem{zlocha2019improving}
M.~Zlocha, Q.~Dou, and B.~Glocker, ``Improving retinanet for ct lesion
  detection with dense masks from weak recist labels,'' in \emph{MICCAI}.\hskip
  1em plus 0.5em minus 0.4em\relax Springer, 2019, pp. 402--410.

\bibitem{li2019mvp}
Z.~Li, S.~Zhang, J.~Zhang, K.~Huang, Y.~Wang, and Y.~Yu, ``Mvp-net: Multi-view
  fpn with position-aware attention for deep universal lesion detection,'' in
  \emph{MICCAI}.\hskip 1em plus 0.5em minus 0.4em\relax Springer, 2019, pp.
  13--21.

\bibitem{mmdetection}
K.~Chen, J.~Wang, J.~Pang, Y.~Cao, Y.~Xiong, X.~Li, S.~Sun, W.~Feng, Z.~Liu,
  J.~Xu, Z.~Zhang \emph{et~al.}, ``{MMDetection}: Open mmlab detection toolbox
  and benchmark,'' \emph{arXiv preprint arXiv:1906.07155}, 2019.

\end{thebibliography}

\appendix

\setcounter{table}{0}
\renewcommand{\thetable}{A\arabic{table}}

\setcounter{figure}{0}
\renewcommand{\thefigure}{A\arabic{figure}}

\section{Ablation Study}

\begin{table}[tb]
	
	\caption{A comparison of ACS variants, with / without pretraining, in terms of \revised{LIDC-IDRI} segmentation Dice, classification AUC, actual memory and runtime speed per iteration. }
	
	\centering
	\begin{tabular*}{\hsize}{@{}@{\extracolsep{\fill}}lcccc@{}}
		\toprule
		& Seg & Cls & Memory (Seg) & Time (Seg) \\
		\midrule
		ACS \textbf{r.} & \textbf{75.1} & \textbf{92.5} & \textbf{6.6 Gb} & \textbf{0.95 s} \\ 
		M-ACS \textbf{r.} & 74.4 & 89.9 & 7.8 Gb & 1.49 s \\ 
		S-ACS \textbf{r.} & 75.0 & 89.3 & 9.9 Gb & 1.58 s\\ 
		\midrule
		ACS \textbf{p.} & \textbf{76.5} & 94.9 & \textbf{6.6 Gb} & \textbf{0.95 s} \\
		M-ACS \textbf{p.} & 75.1 & 92.7 & 7.8 Gb & 1.49 s\\ 
		S-ACS \textbf{p.} & 75.9 & \textbf{95.1} & 9.9 Gb & 1.58 s\\ 
		
		\bottomrule
	\end{tabular*}

	\label{tab:variants-performance}
\end{table}

\subsection{Analysis of ACS Convolution Variants} \label{sec:ablation-variant}
We analyze the variants of ACS convolutions, including Mean-ACS convolutions and Soft-ACS convolutions. We test these three methods on LIDC-IDRI dataset, using the same experiment settings and training strategy. As depicted in Table \ref{tab:variants-performance}, the vanilla ACS outperforms its variants in most situations, and pretraining is useful in all cases. Specifically, Mean-ACS is the worst under pretraining setting, due to its inability to distinguish the view-based difference with a symmetric aggregation. Soft-ACS outperforms others in some cases (\ie, classification with pretraining). \revised{However, it consumes more GPU memory and time at the training stage without significant performance boost. We suspect the key issue of Soft-ACS is the soft weights using Softmax, which tends to be producing high-entropy outputs (\ie, around $1/3$) as Mean-ACS. Nevertheless, it is potential to improve the ACS convolutions by sophisticated optimization techniques (\eg, temperature annealing) to automatically assign the ACS kernel axes}. Memory and time is measured with a batch size of 2, on a single Titan Xp GPU. The memory consuming differs from the theoretical analysis due to PyTorch internal implementation.

\subsection{Whole-Network vs. Encoder-Only Pretraining}

A key advantage of the proposed ACS convolution is that 
it enables flexible whole-network conversion together with the pretrained weights. We thereby validate the superiority of whole-network weight transferring (WN) over encoder-only weight transferring (EO). We train $4$ models in different pretraining setting: entirely randomly-initialized (ACS \textbf{r.}), only the pretrained ResNet-101 backbone (ACS \textbf{p.}EO) on ImageNet (IMN) \cite{deng2009imagenet} and MS-COCO (MSC) \cite{lin2014microsoft}, and whole pretrained model (ACS \textbf{p.}WN) on MS-COCO (MSC) \cite{lin2014microsoft}. The results are shown in Table \ref{tab:encoder-performance}. It is observed that with more pretrained weights loaded, the model achieves better performance (\textbf{p.}WN$>$\textbf{p.}EO$>$\textbf{r.}), and the whole-network pretraining achieves the best. Note that although methods like I3D \cite{carreira2017quo}, Med3D \cite{chen2019med3d} and Video \cite{hara2018can} provide natively 3D pretrained models, apart from the underperforming performance, these pretraining methods are less flexible and versatile than our method. Generally, only the encoders (backbones) are transferred in previous pretraining methods, however the decoders of \emph{state-of-the-art} models are also very large in parameter size, \eg, the DeepLabv3+ \cite{chen2018encoder} decoder (ASPP) represents $27.5\%$ parameters. The previous pretraining methods hardly take care of the scenarios.

\begin{table}[tb]
	\caption{LiTS segmentation performance of ACS DeepLab ``\textbf{r.}" (initialized randomly), ``\textbf{p.}EO-IMN" (encoder-only pretraining on ImageNet \cite{deng2009imagenet}), and ``\textbf{p.}EO-MSC" (encoder-only pretraining on MS-COCO \cite{lin2014microsoft}), ``\textbf{p.}WN" (whole-network pretraining on MS-COCO (MSC) \cite{lin2014microsoft}). The model sizes of pretrained weights out of the whole models are also depicted, parameters from the final random initialized layer are not counted. }
	
	\centering
	\begin{tabular*}{\hsize}{@{}@{\extracolsep{\fill}}lccccc@{}}
		\toprule
		\multirow{2}{*}{Models}& Size of &\multicolumn{2}{c}{Lesion} & \multicolumn{2}{c}{Liver} \\
		& Pretrained Weights & DG &   DPC  &  DG &  DPC \\
		\midrule
		ACS \textbf{r.}& 0 Mb (0\%) & 75.2 & 62.1  & 95.0 & 94.9  \\ 
		ACS \textbf{p.}EO-IMN &170.0 Mb (72.5\%)& 75.3 & 64.3 & 94.7 & 94.0 \\
		ACS \textbf{p.}EO-MSC &170.0 Mb (72.5\%)& 76.1 & 61.6 & 95.5 & 95.0\\
		ACS \textbf{p.}WN &234.5 Mb (100\%) & \textbf{78.9} & \textbf{65.3} & \textbf{96.7} & \textbf{96.2} \\ 
		\bottomrule
	\end{tabular*}

	\label{tab:encoder-performance}
\end{table}

\section{Algorithm and Implementation}

For the sake of completeness, we describe the detailed calculation of ACS convolutions in Algorithm \ref{algo-acs}. Our PyTorch \cite{paszke2017automatic} implementation is open-source at \url{https://github.com/M3DV/ACSConv/}. Using the provided functions, standard 2D CNNs could be converted into ACS CNNs for 3D images with a single line of code, where 2D pretrained weights could be directly loaded. Compared with 2D models, it introduces no additional computation costs, in terms of FLOPs, memory and model size.

\begin{algorithm}
	
	\caption{ACS Convolution}
	\label{algo-acs}
	
	\KwInput{$\boldsymbol{X_i} \in \mathbb{R}^{C_i\times D_i\times H_i\times W_i}$, $\boldsymbol{W}\in 
		\mathbb{R}^{C_o\times C_i\times K\times K}$,\\
		
		padding: $\boldsymbol{p}$, stride: $\boldsymbol{s},$
		dilation: $\boldsymbol{d}$, view : $V=\{a, c, s\}$, \\
		
		kernel split: $(C_o^{(a)},C_o^{(c)},C_o^{(s)})$,
		$\sum_{v}^{V} (C_o^{(v)})=C_o$,\\
		
		\textit{pad}: compute the padded tensor given an axis to satisfy the final output shape same as Conv3D, \\
		
		\textit{unsqueeze}: expand tensor dimension given an axis. 
		
	}
	\KwOutput{$\boldsymbol{X_o} \in \mathbb{R}^{C_o\times D_o\times H_o\times W_o}$}
	
	\nl Compute ACS kernels: $\boldsymbol{W_a}\in \mathbb{R}^{C_o^{(a)}\times C_i\times K\times K\times 1}$,  $\boldsymbol{W_c}\in \mathbb{R}^{C_o^{(c)}\times C_i\times  K\times1\times K}$,
	$\boldsymbol{W_s}\in \mathbb{R}^{C_o^{(s)}\times C_i\times 1\times K\times K}$\:\\
	
	$\boldsymbol{W_a} =$ \textit{unsqueeze} $(\boldsymbol{W}[0:C_o^{(a)}], \mathit{axis}=a)$\;
	
	$\boldsymbol{W_c} =$ \textit{unsqueeze} $(\boldsymbol{W}[C_o^{(a)}:C_o^{(a)}+C_o^{(c)}], \mathit{axis}=c)$\;
	$\boldsymbol{W_s} = $ \textit{unsqueeze} $(\boldsymbol{W}[C_o^{(a)}+C_o^{(c)}:], \mathit{axis}=s)$\;

	\nl Compute view-based 3D features from three views:\\
	
	\For{$v$ in $V=\{a,c,s\}$}
	{ 
		
		
		$\boldsymbol{X_o^{(v)}} = $ Conv3D $($ \textit{pad} $(\boldsymbol{X_i},\boldsymbol{p}, \boldsymbol{s},\boldsymbol{d},\mathit{axis}=v),\boldsymbol{W_v},$
		$\mathit{stride}=\boldsymbol{s}, \mathit{dilation}=\boldsymbol{d})\in \mathbb{R}^{C_o^{(v)}\times D_o\times H_o\times W_o}$\; 
		
	}
	\nl $\boldsymbol{X_o} =$ \textit{concatenate} $([\boldsymbol{X_o^{(a)}},\boldsymbol{X_o^{(c)}},\boldsymbol{X_o^{(s)}}], \mathit{axis}=0)$.
\end{algorithm}

\begin{lstlisting}
from torchvision.models import resnet18
from acsconv.converters import ACSConverter
# model_2d is a standard PyTorch 2D model
model_2d = resnet18(pretrained=True)
# model_3d is dealing with 3D data
model_3d = ACSConverter(model_2d)
\end{lstlisting}

Actual memory consuming and runtime speed are reported in Table \ref{tab:appendix-implementation}. Due to the engineering issues \revised{(PyTorch internal implementation)}, the memory of ACS convolutions is large than that of 2D (2.5D) and 3D convolutions, yet theoretically identical. It is expected to be fixed (6.6 Gb to 5.0 Gb) in further implementation by custom memory checkpointing. Even though time complexity of ACS and 2D convolutions is the same, the parallelism of the ACS convolutions is weaker than that of 2D convolutions. Thereby, the actual runtime speed of ACS convolutions is slower than that of 2D convolutions.

\begin{table}
	
	\caption{Model performance, memory consuming and runtime speed of 2D (2.5D) and 3D and ACS convolutions. }
	\label{tab:appendix-implementation}
	
	\centering
	\begin{tabular*}{\hsize}{@{}@{\extracolsep{\fill}}lcccc@{}}
		\toprule
		& Seg & Cls & Memory (Seg) & Time (Seg) \\
		\midrule
		2D \textbf{r.}  & 68.8 & 89.4 & 5.0 Gb & 0.57s\\ 
		3D \textbf{r.} & 74.7 & 90.3 & 5.0 Gb & 1.01 s \\ 
		ACS \textbf{r.} & 75.1 & 92.5 & 6.6 Gb & 0.95s\\
		\bottomrule
	\end{tabular*}
	
\end{table}

\section{Details of Proof-of-Concept Dataset}

To generate the 2D dataset in the proof-of-concept experiments, we first equally divide a blank $48\times48$ 2D image into four $24\times24$ pieces. We randomly choose $3$ out of the $4$ pieces and in each of the selected piece, we generate a random-size circle or square with same probability at random center. The size is limited in the $24\times24$ piece. Thereby, the generated shape is guaranteed to be non-overlapped. Similarly, for generating 3D dataset, we equally divide a blank $48\times48\times48$ 3D volume into eight $24\times24\times24$ pieces. We randomly choose $4$ out of the $8$ pieces and in each of the selected piece, we generate a random-size cone, pyramid, cube, cylinder or sphere with same probability at random center. The size is limited in the $24\times24\times24$ piece. For both 2D and 3D datasets, we add $\mathcal{N}(0,0.5)$ Gaussian noise on each pixel / voxel. 

\section{More Results on LIDC-IDRI}

Apart from the ResNet \cite{he2016deep} in the main text, we further experiment with the proposed ACS convolutions on LIDC-IDRI lung nodule classification and segmentation task, using VGG \cite{Simonyan15} and DenseNet \cite{huang2017densely}. The experiment settings are exactly the same. As depicted in Table \ref{tab:appendix-vgg} and \ref{tab:appendix-densenet}, the results are consistent with the main text. The 3D (3D and ACS) models outperform the 2D (2.5D) ones. The randomly-initialized ACS models are comparable or better than the 3D models; when pretrained with 2D datasets (\eg, ImageNet \cite{deng2009imagenet}), the ACS models consistently outperform the 3D ones. 

\begin{table}
	\caption{VGG-16 \cite{Simonyan15} results on \revised{LIDC-IDRI} lung nodule segmentation (Dice global) and classification (AUC).}
	
	\centering
	\begin{tabular*}{\hsize}{@{}@{\extracolsep{\fill}}lccccccc@{}}
		\toprule
		Models & Segmentation & Classification \\
		\midrule
		
		2.5D VGG-16 \textbf{r.} & 71.0 & 89.7 \\ 
		2.5D VGG-16 \textbf{p.} & 71.6 & 93.9 \\
		3D VGG-16 \textbf{r.} & 75.0 & 91.7 \\ 
		3D VGG-16 \textbf{p.} I3D \cite{carreira2017quo} & 75.5 & 94.0 \\ 
		\midrule
		ACS VGG-16 \textbf{r.} & 75.2 & 94.2 \\ 
		ACS VGG-16 \textbf{p.}  & \textbf{75.8} & \textbf{94.3} \\ 
		\bottomrule
	\end{tabular*}
	
	\label{tab:appendix-vgg}
\end{table}

\begin{table}
	
	\caption{DenseNet-121 \cite{huang2017densely} results on \revised{LIDC-IDRI} lung nodule segmentation (Dice global) and classification (AUC). }
	
	\centering
	\begin{tabular*}{\hsize}{@{}@{\extracolsep{\fill}}lccccccc@{}}
		\toprule
		Models & Segmentation & Classification \\
		\midrule
		
		2.5D DenseNet-121 \textbf{r.} & 67.4 & 87.4 \\ 
		2.5D DenseNet-121 \textbf{p.} & 71.8 & 92.6 \\
		3D DenseNet-121 \textbf{r.} & 73.6 & 90.0 \\ 
		3D DenseNet-121 \textbf{p.} I3D \cite{carreira2017quo} & 73.6 & 90.0 \\ 
		\midrule
		ACS DenseNet-121 \textbf{r.} & 73.4 & 89.2 \\ 
		ACS DenseNet-121 \textbf{p.}  & \textbf{74.7} & \textbf{92.9} \\ 
		\bottomrule
	\end{tabular*}

	\label{tab:appendix-densenet}
\end{table}

\end{document}